\documentclass[12pt,a4paper]{article}
\usepackage{jheppub} % for details on the use of the package, please
\usepackage{tikz}
\usetikzlibrary{arrows,decorations.pathmorphing,backgrounds,positioning,fit,petri}
\usepackage{bm,amsmath,amssymb,latexsym,graphicx,euscript,multirow,color,url,verbatim}
\usepackage{epsf}
\usepackage{enumerate}
\usepackage{amscd}
\usepackage{amsthm}
\usepackage{epstopdf}
\usepackage{amsmath}
\usepackage{comment}
\usepackage{dsfont}
\usepackage{slashed}
\usepackage{braket}
\usepackage{epsfig}
\usepackage{subfigure}
\usepackage{framed}

\usepackage{chngcntr}
\counterwithin{figure}{section}

\def\half{\frac{1}{2}}
\def\d{\partial}

\def\Tr{\textrm{Tr}}

\def\cS{\mathcal S}

\def\cD{\mathcal D}
\def\cL{\mathcal L}
\def\cO{\mathcal O}
\def\cM{\mathcal M}

\def\half{\frac{1}{2}}
\def\dm{\partial_{\mu}}
\def\dn{\partial_{\nu}}
\def\dmu{\partial^{\mu}}
\def\dnu{\partial^{\nu}}
\def\dz{\partial_z}

\def\Dz{z\dz\frac{1}{z}}
\def\Am{A_{\mu}}
\def\Vm{V_{\mu}}
\def\An{A_{\nu}}
\def\Vn{V_{\nu}}
\def\Amu{A^{\mu}}
\def\Vmu{V^{\mu}}
\def\Anu{A^{\nu}}
\def\Vnu{V^{\nu}}
\def\Az{A_z}

\def\Vz{V_z}

\def\Amz{A_{\mu z}}
\def\Azm{A_{z\mu}}
\def\Vmz{V_{\mu z}}
\def\Vzm{V_{z\mu}}

\def\Anm{A_{\nu\mu}}

\def\Vnm{V_{\nu\mu}}
\def\cd{\cdot}
\def\Ud{U^{\dagger}}

\newcommand{\aat}[1]{a_{#1}}
\newcommand{\vat}[1]{v_{#1}}

\newcommand{\Amat}[1]{A_{\mu}^{(#1)}}
\newcommand{\Anat}[1]{A_{\nu}^{(#1)}}
\newcommand{\Azat}[1]{A_z^{(#1)}}

\newcommand{\Vmat}[1]{V_{\mu}^{(#1)}}
\newcommand{\Vnat}[1]{V_{\nu}^{(#1)}}

\def\nn{\nonumber}

\newcommand{\comments}[1]{}

\title{\boldmath  
On Effective Actions from Holography}

\author[a]{Sophia K. Domokos}
\author[b,c]{Matan Field}

\affiliation[a]{Center for Cosmology and Particle Physics, Department of Physics, New York University, New York, New York 10003, USA}
\affiliation[b]{Department of Physics, Technion, Haifa 32000, Israel}

\affiliation[c]{Department of Mathematics, University
of Haifa at Oranim, Qiryat Tivon 36006, Israel}

\emailAdd{skd5@nyu.edu}
\emailAdd{fmatan@gmail.com}

\abstract{Inspired by holographic Wilsonian renormalization, we propose a novel perspective on the low-energy effective actions of confining gauge theories with gravity duals.  By identifying the IR-boundary value of a certain bulk field as overlapping with the lightest mode of the field theory, we derive its on-shell effective action by integrating over the rest of the geometry. We illustrate the details of this formalism by computing chiral Lagrangian coefficients in a simple AdS/QCD toy model, finding agreement with previous results. At higher orders we obtain new results in that model, including a closed form for the four-pion scattering amplitude to all orders in momentum. 
Finally, we reformulate our method 
in terms of bulk Feynman diagrams.}

\begin{document}
\maketitle
\flushbottom

\section{Introduction}

One can view the Wilsonian approach to quantum field theory
(QFT) as an ordered slicing in the space of fields. Though it is the full
path integral which ultimately determines the values of physical observables,  we are
free to foliate this space into co-dimension one
slices, which we incrementally integrate out to produce a
renormalization group (RG) flow. For a given foliation, we can, for instance, associate a particular length scale $l$ with each
of the field theory modes ($M$), and integrate over modes that
are assigned  lengths $l<\delta$ to define a reduced field space
and an effective action $S_{\delta}$:
\begin{align}\label{WRG}
Z_{QFT}=\int \cD M_{l\leq\delta}\cD M_{l>\delta}e^{-S_{0}[M_<,M_>]}\equiv\int\cD M_{l>\delta}e^{-S_{\delta}[M_>]}~.
\end{align}
Formally, the bare action $S_{0}$ is defined with a regulator $\epsilon$. 
There are countless ways to foliate field space, and to assign each mode a length (or energy) scale. Different foliations define different RG flows.
A sharp cutoff in Euclidean particle momenta ($k_E^2<\Lambda^2$)  is a common renormalization scheme. 
However, one can in principle construct  any other arbitrarily intricate 
slicing.\footnote{We might require a `good' slicing to lead to a \textit{local} effective action along the flow. } 

An intriguing new renormalization scheme arose in the context of AdS/CFT: 
\textit{holographic Wilsonian renormalization} (HWR),  defined for
gauge theories having asymptotically AdS (aAdS) classical gravity duals. The new scheme was inspired 
by the identification of the radial position in the bulk with energy scale in the gauge theory, $z\sim 1/\Lambda$. 
Evolution along the bulk radial direction thus provides a geometrical picture for Wilsonian RG flow in the field theory. 
Initially \cite{Akhmedov:1998vf,deBoer:2000cz,Bianchi:2001de}, the holographic RG approach related the RG equations to 
Hamilton-Jacobi equations in the bulk, with the radial direction playing the role of time. This treatment was not Wilsonian in 
nature, as the flow depended on information at the IR (such as bulk regularity).
More recently,  \cite{Heemskerk:2010hk, Faulkner:2010jy} formulated
 a truly Wilsonian holographic renormalization, with a partition of bulk modes according to their radial position. (See also the earlier attempt of \cite{Balasubramanian:1999jd} and a recent extension of these ideas in \cite{Balasubramanian:2012hb}.)

In the HWR formalism of \cite{Heemskerk:2010hk, Faulkner:2010jy}, the separation of the path integral \eqref{WRG} into modes 
$M_>$ and $M_<$ is identified with a separation of the bulk path integral into integrations
of bulk fields ($\Phi$) above and below a radial value,
\begin{align}\label{eq:bulk_separation}
Z_{bulk}= \int \cD\Phi_{z\leq l}\cD\Phi_{z>l}e^{-\cS_{0}[\Phi_<,\Phi_>]}\equiv \int \cD \Phi_{z>l}e^{-\cS_{l}[\Phi_>]}~.
\end{align}
 $\cS_{0}$ includes the original bulk action (regularized at $z=\epsilon$), and possibly also an action at the UV boundary. $\cS_{l}$ is the total effective action. It includes the original bulk action on the remaining slice $z>l$,  
 and an induced boundary action on the new cutoff surface at $z=l$. The new UV boundary action is related to the induced Wilsonian effective action at scale $\delta\sim l$.\footnote{The precise relation between the scales is ambiguous and depends on the slicing definition.}

Despite its  intuitive appeal, the HWR formalism's most crucial ingredient remains obscure:  what is the precise renormalization scheme in QFT that corresponds to the radial cutoff in the bulk?\footnote{See \cite{Gubser:1999vj,Verlinde:1999fy,ArkaniHamed:2000ds,Rattazzi:2000hs,PerezVictoria:2001pa,Brattan:2011my} for related works. }
This question is related to a detailed understanding of the holographic duality: the precise mapping of local 
bulk excitations to boundary modes and the ``emergence" of the radial direction. We do not address these fundamental questions here. 
Instead we solve a simpler problem that we hope may shed  new light on the subject.

In this work we explain how to compute low-energy effective actions (LEEAs) for strongly-coupled confining gauge theories with  gravity duals. The framework we propose is driven by the HWR formalism, which for this particular task can be made completely well-defined. Concretely, our HWR-like procedure yields a scheme-independent, on-shell 
 effective action, which is computed in a more direct and efficient way as compared to previous techniques.

We consider the large $N$ limit of confining gauge theories in $d$-dimensional flat space, which 
have a classical gravity dual in an asymptotically $AdS_{d+1}$ space ($aAdS_{d+1}$). 
   For the sake of simplicity we use a ``hard-wall"  model, in which confinement is induced by sharply cutting off the bulk geometry at a finite radial value, $z<L$ \cite{DaRold:2005zs,Erlich:2005qh}. 
   We also assume the gauge theory to be IR free but non-trivial, so there is some interesting 
weakly-coupled description of the physics at the IR. When the $(d+1)$-dimensional bulk theory terminates on a 
non-degenerate $d$-dimensional surface at $z=L$, we can perform the whole path integration along the $z$-direction in the spirit of HWR (from UV to IR), and reach an effective action in $d$ dimensions. According to the HWR prescription it is natural to identify the resulting action as the LEEA for the lightest mode.   
   While the effective action at finite cutoff is scheme-dependent (and, as mentioned above, we know nothing about the scheme in the HWR formalism), we might expect the LEEA obtained from integrating over the whole geometry to be somewhat universal.
   To be more precise, the complete integration of bulk geometry should exclude a mode that is localized on the IR boundary. This mode is not integrated out, and the effective action we compute depicts its dynamics. Indeed, as we explain, the choice of IR mode is to some extent arbitrary, as it affects only off-shell data. 
The on-shell action, or the S-matrix, is invariant under general field redefinitions, and as such will be insensitive to the precise choice of the mode that is left unintegrated, as long as it has `anything to do' with the true light degree of freedom  we are after, in a way that we will make more precise below.

While some of the perspective and formalism  we present here is novel, as is the methodology,
many of the ideas we describe have been floating around in the literature for some time. For example, in the
context of holographic hydrodynamics, \cite{Nickel:2010pr} presented closely-related work (see also the review of \cite{Banerjee:2011tg}), giving 
an effective description for the long wavelength behavior of
holographic fluids. 
Other holographic models have been used to  derive effective actions in a variety of contexts, 
such as \cite{Hoyos:2012xc,Hoyos:2013gma,Bajc:2013wha}, which study the low-energy dynamics of
 the Goldstone boson of broken conformal invariance (the dilaton).
\cite{McGarrie:2012fi} describes a holographic framework for spontaneous SUSY breaking,
 deriving the 4d effective
action for Goldstinos  by integrating out bulk fields. In a more phenomenological context, a very closely related work is that of \cite{Panico:2007qd} (see also the earlier papers of 
\cite{Luty:2003vm, Barbieri:2003pr}), defining a similar effective action, though the procedure and perspective are quite different from ours. Other related phenomenological works with a  holographic bent include \cite{Lewandowski:2002rf} and subsequent works. 
Finally, we will illustrate the details of our methodology on a well-known AdS/QCD example, the Hirn-Sanz model \cite{Hirn:2005nr}, in which
results similar to ours have been derived using other techniques. The pion effective action at four-derivative order was computed
already in the original papers \cite{DaRold:2005zs}\cite{Erlich:2005qh}; the six-derivative result was recently presented in  \cite{Colangelo:2012ipa}.
Using our techniques, however, we will extend these results to infinite orders in derivatives for the cases of four and six external 
pions (the former to be written in a closed form).

 The paper is organized as follows. In section \ref{sec:HIREA} we discuss low-energy effective actions (LEEAs) of confining gauge 
theories with a classical gravity dual. We first review  the ``traditional'' Kaluza-Klein method of computing the LEEA 
holographically, then describe in greater detail our prescription for the holographic Wilsonian LEEA. We give 
arguments for the robustness of the procedure, and make contact with the HWR formalism. In section \ref{sec:example} we work out an example that illustrates the simplicity of our prescription: computing the on-shell LEEA for the pions of a simple AdS/QCD model. We compare to known results in the leading orders of the momentum expansion, and extend them further, in some cases to all orders. In section \ref{sec:diagrammatics} we reformulate the procedure in terms of Feynman diagrams in the bulk that are sourced by our pion field, defined on the IR boundary. This provides some additional intuition for the HWR process and simplifies the computations significantly. We conclude in section \ref{sec:discussion} with a brief summary and some interesting open questions. Technical details are deferred to the Appendices.

\section{Low-Energy Effective Actions from Holography}\label{sec:HIREA}

Let us first review the standard method for deriving LEEAs from holography via KK reduction. Throughout this paper we refer to a general $d$-dimensional, strongly-coupled confining gauge theory at large $N$, that admits a $(d+1)$-dimensional gravity dual. We consider an aAdS bulk geometry with radial coordinate $z$ and a general warp factor,\begin{align}
ds^2 = %g_{MN} dx^M dx^N =
w^2(z)\left(\eta_{\mu\nu}dx^\mu dx^\nu -dz^2\right)~,
\end{align}
that obtains a conformal boundary, $w(z)\sim1/z$, at small $z$.\footnote{We make the simplifying assumption that the extra compact manifold is in a direct product with the AdS part, in which case it plays no special role in what follows and we can simply ignore it. We use capital Latin letters $M=(\mu,z)$ to denote 5d coordinates, and Greek indices to denote flat space directions. We use the mainly minus convention, and 5d (4d) indices are lowered with $g_{MN}$ ($\eta_{\mu\nu}$).}
We consider confinement induced by sharply cutting off space at finite radial coordinate $z<L$, leaving the generalization to smooth confining geometries for future work.

\subsection{Holographic Low-Energy Effective Action via Kaluza-Klein}\label{subsec:KKIREA}
Gauge-invariant excitations, such as mesons and glueballs, are encoded holographically as normalizable modes of bulk fields. Each bulk field, upon KK reducing along the radial direction, gives rise to an infinite tower of such excitations, all having the same quantum numbers in the $d$-dimensional theory.   For example, a single vector gauge field in five dimensions produces a tower of vector mesons in four dimensions. Let's consider, for concreteness, a bulk scalar field with action
\begin{align}\label{eq:generalaction}
S=\int d^{d+1}x \sqrt{G}
\left[\frac{1}{2}\d_M\Phi\d^M\Phi-\frac{1}{2}M^2\Phi^2-\frac{1}{4!}\lambda \Phi^4\right]~,
\end{align}
and equations of motion
\begin{align}
\left(\d^2-w^{-d+1}\dz w^{d-1}\dz+M^2w^2\right)\Phi=-\frac{1}{3!}\lambda w^2\Phi^3~.
\end{align}
We can use the standard expansion, 
 \begin{align}\label{eq:phiexpn}
\Phi(x,z) &= \sum\limits_{n=1}^\infty\phi_n(x)\psi_n(z)~,
 \end{align}
with the eigenfunctions of the radial quadratic operator (and appropriate boundary conditions inherited from those of $\Phi$),
 \begin{align}
 \left(w^{-d+1}\dz w^{d-1}\dz-M^2w^2\right)\psi_n=m_n^2\psi_n~.
\end{align}
Plugging the expansion back into the action and explicitly performing the $z$-integration, one finds an infinite tower of interacting $d$-dimensional fields with masses $m_n$,
\begin{align}
S_{KK}
=&\int d^dx \left[   \sum\limits_n \half\dm\phi_n\dmu \phi_n -\half m_n^2\phi_n^2 -\lambda\sum\limits_{n_1\leq n_2\leq n_3\leq n_4} v_{n_1,n_2,n_3,n_4} \int d^dx \phi_{n_1} \phi_{n_2} \phi_{n_3} \phi_{n_4}  \right]~.
\end{align}
Momenta in the radial direction translate to $d$-dimensional masses, and the spectrum is discrete due to the effectively-finite size of the radial direction. 
Relative couplings between $d$-dimensional resonances are given by overlap integrals of the corresponding $z$-momentum wave functions,
\begin{align}
v_{n_1,n_2,n_3,n_4} = \int_0^{L} dz w^{d+1}\psi_{n_1}\psi_{n_2}\psi_{n_3}\psi_{n_4}~.
\end{align}
We now have a $d$-dimensional theory of an infinite tower of resonances; this is the \textit{dual resonance model}. If we are primarily interested in low-energy physics, say, in terms of the lightest field $\phi_1$, we may proceed with the standard ($d$-dimensional) procedure:    integrating out all heavy resonances with masses greater than $m_1$,
\begin{align}
Z=\int \cD\Phi ~ e^{iS^{(d+1)}[\Phi]} =\left(\prod\limits_{n} \int \cD\phi_n\right) e^{iS^{(d)}[\{\phi_n\}]}\equiv\int\cD\phi_1 e^{iS^{(d)}_{eff}[\phi_1]}~.
\end{align}
Note that in the above we have completely integrated out all massive fields, and none of the modes of the lightest one $\phi_1$, for which we obtain the \textit{exact} effective action. At large $N$ the bulk is at tree level, so the resulting action is also a 1PI effective action. 
While this is not the same as the (deep-IR) Wilsonian effective action, the two should agree when evaluated on-shell. 
The effective action is expected to be local only below $m_2$, where a `good' derivative expansion applies, with higher-derivatives terms scaled with powers of $p/m_2$. Thus, it is mostly useful when $m_2$ is sufficiently gapped from $m_1$. However, when it is possible to compute a process to all orders in derivatives and sum it up, the result will be valid at all energies. 
That is indeed what we will encounter below, for the four-pion scattering amplitude.

\subsection{A Holographic Wilsonian Approach to IR Effective Actions}\label{subsec:HWIREA}

In the KK procedure we split up the bulk fields into modes according to their radial \textit{momenta},  integrating out all but the lightest mode to get the LEEA. 
In some sense, this procedure does not take full advantage of the holographic description: the information is translated to a $d$-dimensional language ``too early.''
In the HWR formalism one partitions the field into modes according to their radial \textit{position}.
We propose that one can retrieve the same physical information at the IR by integrating out the whole bulk geometry apart from some modes on the IR boundary.
 Keeping manifest the holographic radial dimension both simplifies the computation and adds some geometrical intuition. 
Concretely, one can take the ``IR  limit'' of the formulation in \eqref{eq:bulk_separation}, by sliding the cutting surface at $z=l$ all the way down to where the bulk terminates, $l\rightarrow L$:
\begin{align}\label{eq:our_separation}
Z_{bulk}= \int \cD\varphi \int_{\begin{smallmatrix}\Phi(L)=\varphi\end{smallmatrix}} \cD\Phi_{z<L}e^{-\cS_{0}[\Phi]}\equiv\int \cD \varphi e^{-\cS_{L}[\varphi]}~.
\end{align}
In this limit we are left with an effective action $\cS_{L}[\varphi]$ for the unintegrated $d$-dimensional boundary field,
\begin{align}\label{eq:our_assignment}
\varphi(x)\equiv \Phi(x,L)~.
\end{align}
In the spirit of HWR it is natural to identify $\varphi$ with a low-energy mode, and $\cS_{L}[\varphi]$ with its LEEA. We now examine this idea more closely.

\textbf{Which effective action?}
  Since $\varphi$ is not being integrated over in \eqref{eq:our_separation}, one might naively guess that $\cS_{L}[\varphi]$ is an off-shell effective action.
This is not the case, however. Even before imposing \eqref{eq:our_assignment}, the path integral in \eqref{eq:our_separation} already contains the original (two) boundary conditions, and together with the additional assignment of \eqref{eq:our_assignment} it is classically over-constrained.\footnote{However, we assume the assignment of \eqref{eq:our_assignment} to be algebraically consistent with the IR boundary condition, otherwise we simply change our IR-mode identification. See below for further details.} It is consistent only if the assignment of $\varphi$ sits on the classical saddle point of the path-integral with the original boundary conditions, which then translates to $\varphi$ being a solution to its own equation of motion. Therefore, we identify $\cS_{L}[\varphi]$ with the \textit{on-shell} effective action of $\varphi$. 

In Lorentzian signature, one also needs to impose initial conditions when solving the bulk equations of motion. In the effective theory, these amount to specifying the heavy on-shell resonances excited above the vacuum, which correspond to normalizable modes in the bulk. Because we are interested in the $\varphi$ effective action in the vacuum, we require that the solutions for all fields -- expressed exclusively in terms of $\varphi$ -- vanish when $\varphi\rightarrow 0$.  

Finally, we should ask: to what extent is $\cS_{L}$ a low-energy action? It is the effective action that has already accounted for all fluctuations of all fields apart from the $\varphi$ mode. It is then a low-energy action to the extent that $\varphi$ is a low-energy mode. 

\textbf{Effective action of what?}
Although we have identified the on-shell effective action of an IR boundary mode $\varphi$, eventually we are interested in the on-shell LEEA of the lightest mode in the gauge theory, $\phi_1$, which is the only propagating mode at energies just above $m_1$. The two modes, $\varphi$ and $\phi_1$, are not identical, nor are their off-shell effective actions. Their on-shell effective actions, however, might be the same. This is due to  the ``universality of the S-matrix'', a theorem stating that the S-matrix of arbitrary scalars is invariant under  field redefinitions of the form
\begin{align}\label{eq:overlap}
\phi'=c_1\phi+c_2\phi^2+...
\end{align}
This holds even if $\phi,\phi'$ denote multiple fields (and the $c_i$ matrices), or if the $c_i$'s depend on momentum.\footnote{The exact statement is that the scattering amplitudes of $\phi'$ particles with Lagrangian $\cL_{\phi'}$ equal the corresponding scattering amplitudes of $\phi$ particles with Lagrangian $\cL_{\phi'(\phi)}$. It stems directly from a simple pole analysis (see for example\cite{Donoghue:1992dd,Bando:1987br}).} A similar statement holds for the on-shell effective actions of $\phi$ and $\phi'$, in which  the overall normalization $c_1$ needs to be accounted for separately.  
For our purposes, we simply conclude that as long as we choose the IR mode $\varphi$ so that it overlaps with the lightest mode $\phi_1$ in the way defined by \eqref{eq:overlap}, the off-shell effective actions of the two modes may differ, but their on-shell actions will agree.  We can then identify
\begin{align}\label{eq:prescription2}
e^{-\cS^{(on-shell)}_{IREA}[\phi_1]}=\int_{\begin{smallmatrix}\Phi(L)=\phi_1\end{smallmatrix}} \cD\Phi_{z<L}e^{-\cS_{0}[\Phi]}~.
\end{align}
More generally, in addition to its overlap with $\phi_1$, $\varphi$ can also mix with heavier fields. We can still study $\phi_1$'s  scattering amplitudes using those of $\varphi$ in such cases; if we put all incoming $\varphi$ states on the mass-shell of $\phi_1$, we  isolate the  leading singularity that corresponds to the $\phi_1$'s S-matrix.
There will often be some freedom in our choice of which IR mode to leave unintegrated. Some options may simplify computations considerably,
but roughly speaking any choice of bulk mode that has some coupling with the lightest mode of the gauge theory will work.  

In a general holographic setup, the IR mode $\varphi$ will mix with the whole tower of KK excitations $\phi_n$, but there exists a simpler case where it can be set to mix only with the lightest one. This happens when the lightest mode is a massless Nambu-Goldstone boson (NGB), which arises from a spontaneously broken symmetry in the gauge theory. We will explore this example in section \ref{sec:example}.
Equation \eqref{eq:prescription2} is our main conceptual output, and we now explain how to use it in broad strokes. When we work out a specific example  in section \ref{sec:example}, we
will explain the technical details of the procedure.

\textbf{The prescription in practice:}  To compute the effective action in \eqref{eq:prescription2} we need to: 
(a) solve the equations of motion of $\Phi$ in terms of $\varphi$ 
(with the  assignment \eqref{eq:our_assignment} in addition to the original two boundary conditions), then (b) plug the solution back into the original action and integrate over the radial coordinate. 
As explained earlier, requiring the existence of a solution for $\Phi$ (in terms of $\varphi$), demands that $\varphi$  satisfy a constraining equation, which then corresponds to its equation of motion in the 
effective theory. Note that this off-shell information is \textit{complementary} to the obtained on-shell effective action.

From now on we concentrate on the LEEAs of massless particles, $\phi_0$,  which we can probe at arbitrarily low energies. The effective action then has a useful derivative expansion, as derivatives scale with the low-energy momentum transfer $p/m_1$, where $m_1$ is the lowest mass that was integrated out. We also assume the massless field itself to scale with $p/m_1$, so that we can also expand order by order in the IR mode itself. This generally holds true, on dimensional grounds, if the theory has a single dimensionful scale and no dimensionless parameters, as for example in the case of QCD.
 The prescription is simple:
\begin{itemize}
\item First, expand the equations of motion of the various fields order by order in $\varphi$.
Note that $\Phi(x,z)$, the bulk field of which $\varphi$ is an IR value, receives a non-zero contribution at leading order in  $\varphi$. 
Other fields, denote them $\Psi(x,z)$, are sourced by $\varphi$ through the equations of motion, and have their first non-zero coefficient at $O(\varphi^2)$ if their mixed two-point function with $\Phi$ vanishes, $\langle \Psi\Phi\rangle=0$. Note that we compute $\varphi$'s LEEA \textit{in the vacuum}, i.e. we excite no heavy on-shell particles, and thus the solution is required to vanish when taking $\varphi\rightarrow 0$. (This condition is crucial to singling out a unique solution to the equations of motion.)  In particular, no field in our solution would have a term at $O(\varphi^0)$.
\item Solve the equations order by order in $\varphi$. At each order we will encounter free equations of motion, with a source term obtained from lower-order solutions.
\item Finally, plug the solution into the action and integrate over $z$ to obtain the on-shell effective action for $\varphi$.
\end{itemize}
Note that in \eqref{eq:our_separation} we have kept the UV and IR boundary conditions on bulk fields implicit.
One might also be concerned that $\varphi(x)\equiv\Phi(x,L)$ explicitly conflicts with the IR boundary condition, e.g. if it is Dirichlet, $\Phi(x,L)=0$. Thanks to the universality of the S-matrix described above, we have the freedom in such a case to make another choice for our boundary field, e.g. $\varphi(x)\equiv\d_z\Phi|_{z=L}$. Whatever the IR boundary conditions, we simply choose $\varphi$ in a way that does not conflict with the boundary conditions on bulk fields.

In section \ref{sec:diagrammatics} we describe an equivalent procedure in terms of bulk Feynman diagrams which allows one to compute $\cS_{on-shell}[\varphi]$ more directly.
First, however, we demonstrate a pedestrian version of the procedure on a concrete example.

\section{An Example: AdS/QCD}\label{sec:example}

In order to illustrate our prescription, we now work
out the IR effective action for the massless modes of a well-studied example: hard wall AdS/QCD. We emphasize that our goal here is not to discuss the efficacy of this specific model as a dual for QCD (and thus we exclude any motivations for it), but rather to demonstrate the details of our prescription on a concrete example. Towards the end of the section we will reproduce known results for this model, and then extend them to higher -- in some cases, infinite -- derivative order.

\subsection{The Model of Hirn and Sanz}\label{subsec:HSreview}
We briefly review the AdS/QCD model of Hirn and Sanz \cite{Hirn:2005nr}.
The bulk geometry is simply a slice of AdS$_5$, in Poincar\'{e} coordinates,
with an IR boundary at $z=L$ and the usual UV conformal boundary at $z=0$.\footnote{UV divergences conventionally require one to impose a holographic UV regulator, putting the boundary at $z=\epsilon$, eventually taking the  conformal limit at the end of the computation, $\epsilon\rightarrow 0$. Our analysis is insensitive to those divergences and we can take the UV boundary to $z=0$ immediately.}
 The metric is
\begin{align}
ds^2=\left(\frac{R}{z}\right)^2\big( \eta_{\mu\nu}dx^\mu
dx^\nu-dz^2 \big)~, %~~\quad z\in\left[0,L\right]~,
\end{align}
  where $R$ is the AdS radius. We ignore metric fluctuations.
 The global flavor symmetry currents of (massless) QCD are dual to bulk gauge fields of \makebox{$SU(N_f)_L\times SU(N_f)_R$}, labeled $L_M(x,z),R_M(x,z)$, with field strengths 
\begin{align}
&L_{MN}=\d_ML_N-\d_NL_M-i\big[L_M,L_N\big]\nn\\
&R_{MN}=\d_MR_N-\d_NR_M-i\big[R_M,R_N\big]~,
\end{align}
and Yang-Mills action,\footnote{We ignore the Chern-Simons term which would be irrelevant at our working order.}
\begin{align}
\cS_{5d}=-\frac{1}{4g_5^2}\int d^5x\sqrt{g} \ \Tr\Big\{ L_{MN}
L^{MN}+R_{MN}R^{MN} \Big\}~.
\end{align}
Under 5d gauge transformations, with gauge group elements $L(x,z)$ and
$R(x,z)$, 
 the gauge fields transform as usual,
$L_M\equiv L_M^a\frac{T^a}{\sqrt{2}}\rightarrow LL_ML^{\dagger}+iL\d_M L^{\dagger},~
$
 with the generators of $SU(N_f)$ normalized by $\Tr(T^aT^b)=2\delta^{ab}$.
It is natural to work with the vector and axial gauge fields
\begin{align}
 V_M=\half\big(L_M+R_M\big)~,~~~A_M=\half\big(L_M-R_M\big)~,
 \end{align}
and with the corresponding field strengths
\begin{align}\label{eq:VMN_AMN}
&V_{MN}=\d_{[M}V_{N]}-i\big[V_M,V_N\big]-i\big[A_M,A_N\big]~,\nn\\
&A_{MN}=\d_{[M}A_{N]}-i\big[V_M,A_N\big]-i\big[A_M,V_N\big]~,
\end{align}
with which the bulk action reads
\begin{align}\label{eq:HS_action2}
\cS_{5d}=-\frac{1}{2g_5^2}\int d^5x\sqrt{g} \ \Tr\Big\{ V_{MN}
V^{MN}+A_{MN}A^{MN} \Big\}~.
\end{align}
We will consider only normalizable modes (i.e. no background sources in the gauge theory), and impose vanishing UV boundary conditions (BCs),\footnote{The generalization to include external sources is straightforward.}
\begin{align}\label{eq:BCs_UV}
V_{\mu}^a(x,0)=A_{\mu}^a(x,0)=0~.
\end{align}
At the IR, (non-)gauge-invariant BCs are imposed on the (axial-)vector gauge field
\begin{align}\label{eq:BCs_IR}
& V_{\mu z}^a(x,L)=A_{\mu}^a(x,L)=0~,
\end{align}
in order to realize chiral symmetry breaking.
This is the entire field content of the model, minimally capturing the global symmetries of (massless) QCD and their breaking pattern.

We will work in the gauge
\begin{align}\label{eq:gauge_fixing}
V_{z}=0~,
\end{align}
which is compatible with the boundary conditions.
 We cannot fix a similar gauge for the axial gauge field, due to the non-trivial gauge holonomy along the radial direction (see \eqref{eq:holonomy}), induced by the symmetry-breaking BCs. The closest we can achieve might be taking
 \begin{align}\label{eq:AzGF}
 \Az=z f(x)~,
 \end{align}
for some arbitrary function $f(x)$. For the time being we will keep $\Az$ general, and later on we will see how this gauge can be consistently fixed \textit{on-shell} (i.e. under the equations of motion).

From \eqref{eq:HS_action2}, the equations of motion for the gauge fields are
\begin{align}\label{eq:EOM}
 &\frac{1}{\sqrt{g}}\d_M\sqrt{g}V^{MN}=i\big[V_M,V^{MN}\big]+i\big[A_M,A^{MN}\big]~,\nn\\
 &\frac{1}{\sqrt{g}}\d_M\sqrt{g}A^{MN}=i\big[V_M,A^{MN}\big]+i\big[A_M,V^{MN}\big]~.
\end{align}
In Appendix \ref{app:equationsofmotion} we provide them in components. Ultimately we will be interested in the on-shell action, so it is useful to already plug \eqref{eq:EOM} into \eqref{eq:HS_action2} and get an on-shell bulk action
\begin{align}\label{eq:bulk_action_on-shell}
\cS_{5d}^{(on-shell)}= \frac{-i}{2g_5^2}\int d^5x ~
\sqrt{g}\Tr\bigg\{\Big(\big[V_M,V_N\big]+\big[A_M,A_N\big]\Big)V^{MN}+2\big[V_M,A_N\big]A^{MN}\bigg\}~,
\end{align}
where the boundary terms vanish with our BCs \eqref{eq:BCs_UV}\eqref{eq:BCs_IR}. 

Finally, note that from here on we set $L=1$ and restore it at the end of each computation by dimensional analysis. Since the coupling $g_5^2$ has dimensions of length, 
it is convenient to define $g_5^2\equiv R g_4^2$. The dimensionless coupling $g_4^2$ is the effective coupling in the four-dimensional theory, which at the IR eventually relates to the pion decay constant $g_4^2\sim 1/f_{\pi}$, as we will see below.

\subsection{Applying the Prescription}\label{subsec:prescription}
Chiral symmetry is spontaneously broken in the gauge theory, and accordingly we expect to find a weakly interacting theory of  massless pions. Using the prescription outlined above, we can compute 
the low-energy effective action of the pions, the (on-shell)  chiral Lagrangian.
First, we need to identify a mode that has an overlap with the pion. As the chiral symmetry breaking is embedded in the bulk axial gauge field, it is natural to identify the pion
with the IR boundary value of the axial mode,\footnote{This assignment is not gauge-invariant, even when restricting to gauge transformation that respect the BCs \eqref{eq:BCs_UV}\eqref{eq:BCs_IR} and gauge fixing \eqref{eq:gauge_fixing}. Gauge freedom that modifies the boundary value of $\Az$ can either be fixed, or will remain as a gauge redundancy in the effective theory of $\pi(x)$.}
\begin{align}\label{eq:our_pion}
\pi(x)\equiv A_z(x,1)~.
\end{align}
We dub this the \textit{holographic pion}. It overlaps with the ``Nambu-Goldstone pion'' (\textit{NG pion}), defined via its transformation under the broken flavor symmetry as in \eqref{eq:U_trans_law}. Our identification of the pion is certainly not unique.\footnote{For example, we could have also defined $\pi=\dz\Az|_{z=L}$, or even impose a non-linear relation.} 
We will see below that there is some gauge freedom left (after the symmetry-breaking BCs and gauge fixing of $\Vz$) that allows us to ``reshuffle" the pion between the boundary values of $\Az$ and $\Am$. Thus  we can prevent the pion from overlapping with $\Am(x,1)$ at leading order (so that $\Am=O(\pi^3)$), which will prove quite useful.  We will also see that the same gauge freedom can be used order by order in $\pi(x)$ to set a complete gauge fixing -- on-shell, under the equations of motion -- of the form \eqref{eq:AzGF}. To all orders we thus have 
\begin{align}\label{eq:Az_gauge_fix}
\Az(x,z)=z\pi(x)~.
\end{align}
In this simple model the NG pion itself is easily identified: it is the gauge holonomy $\pi\sim\int_0^1 dz \Az$, which agrees with our choice up to a normalization factor. We return to this later, but for the purpose of demonstrating our formalism we feign ignorance of this point, and go on working with $\pi(x)$ as defined in \eqref{eq:our_pion}.
With this definition the pion serves as an IR source for bulk fields. We now  solve for these fields in terms of $\pi(x)$.

\textbf{Leading order solution:} By parity, we see  that  $\Vm$ can only contain even powers of $\pi$, while $\Az,\Am$ contain only odd powers.
At leading order $O(\pi)$, we have the free equations of motion,
\begin{subequations}\label{eq:eom(1)}
\begin{align}
\dmu\big(\dm\Azat{1}-\dz\Amat{1}\big)&=0 ~, \label{eq:eom(1a)} \\
z\dz\frac{1}{z}\left(\dm\Azat{1}-\dz\Amat{1}\right)-\dnu\left(\dm\Anat{1}-\dn\Amat{1}\right)&=0~, \label{eq:eom(1b)}
\end{align}
\end{subequations}
where the superscript denotes the order in $\pi$. At this point we need to choose the physical state we want to perturb around. As explained in section \ref{sec:HIREA}, computing the pion effective action in the vacuum means demanding that the solution vanish when taking $\pi\rightarrow 0$; this implies that \textit{all fields must be order $O(\pi)$ or higher}. The general solution  to \eqref{eq:eom(1)} is, for generic $\pi(x)$, 
 \begin{align}\label{eq:solution(1)_wrong}
&\Azat{1}(x,z)=a'(z)\pi(x)~, \nn\\
&\Amat{1}(x,z)=a(z)\dm \pi(x)~.
\end{align}
At this order \eqref{eq:solution(1)_wrong} is pure gauge, and it can be shown that for an unconstrained $\pi(x)$ the solution remains pure gauge also at higher orders, and so is trivial. Thus, for a non-trivial solution we are forced to consider functions $\pi(x)$ that obey some constraining equation. Allowing for
\begin{align}\label{eq:pi_eom(1)}
\d^2\pi(x)=O(\pi^3)~,
\end{align}
the general solution is now of the form
\begin{align}\label{eq:solution(1)}
&\Azat{1}(x,z)=\big(a'(z)+cz\big)\pi(x)~, \nn\\
&\Amat{1}(x,z)=a(z)\dm \pi(x)~.
\end{align}
Exhausting our gauge freedom and applying the  BCs, we find
\begin{align}\label{eq:solution(1)final}
\Azat{1}(x,z)=z\pi(x)~,~~\Amat{1}=0~.
\end{align}
As explained earlier, \eqref{eq:pi_eom(1)} is to be identified with the pion's leading-order equation of motion in the effective theory. Note that \eqref{eq:solution(1)final} is always a solution to the $\Am$ equation of motion \eqref{eq:eom(1b)} at this order, and that it becomes also a solution to the $\Az$ equation \eqref{eq:eom(1a)} with the pion's equation \eqref{eq:pi_eom(1)}. This rule recurs at higher orders.

\textbf{Second order solution:} Now consider the equations of motion \eqref{eq:eom} to second order in $\pi(x)$,
\begin{align}\label{eq:eom(2)}
\dz\left(\d\cdot V^{(2)}\right)&=0~,\nn\\
z\d_z\frac{1}{z}\dz\Vmat{2}-\dnu\left(\dn\Vmat{2}-\dm\Vnat{2}\right)&=
-i\big[\Azat{1},\dm\Azat{1}\big]~.
\end{align}
Considering only normalizable modes and plugging in the first order solution \eqref{eq:solution(1)final}, these become
\begin{subequations}\label{eq:eom(2)}
\begin{align}
\d\cdot V^{(2)}&=0~, \label{eq:eom(2a)} \\
z\d_z\frac{1}{z}\dz\Vmat{2}-\d^2\Vmat{2}&= -iz^2\big[\pi,\dm\pi\big]~.\label{eq:eom(2b)}
\end{align}
\end{subequations}
In Appendix \ref{app:all-order solution} we solve \eqref{eq:eom(2)} with separation of variables and find
\begin{align}\label{eq:solution(2)}
\Vmat{2}(x,z)%&=z^2\int\frac{d^4 k}{(2\pi)^4}\int\frac{d^4k'}%{(2\pi)^4}k'_{\mu}\big[\pi(k-k'),%\pi(k')\big]\frac{1}{k^2}\bigg(1-\frac{2}{k z}\frac{J_1(kz)}{J_0(k)}\bigg)e^{ik\cd x}~\nn\\
&= \frac{iz^2}{\d^2}\left(1-\frac{2}{z\d}\frac{J_1(z\d)}{J_0(\d)}\right)\big[\pi(x),\dm\pi(x)\big] \nn\\
%&= i\vat{2,1}(z)[\pi,\dm\pi] + i\vat{2,3}(z)\Big[\dnu\pi,\dn\dm\pi\Big] \nn\\
&=-iz^2\bigg(\frac{1}{8}\left(z^2-2\right)+\frac{1}{192}\left(z^2-3\right)^2\d^2+O(\d^4)\bigg)\big[\pi,\dm\pi\big]~,\nn\\
%\vat{2,1}&=-\frac{1}{8}z^2(z^2-2)~,\nn\\
%\vat{2,3}&=-\frac{1}{96}z^2(z^2-3)^2~.
\end{align}
where the $J$'s are Bessel functions and $\d\equiv\sqrt{\d^2}$.
Note again that \eqref{eq:solution(2)} is the unique solution to \eqref{eq:eom(2b)} (with BCs (\ref{eq:BCs_UV}-\ref{eq:BCs_IR})), and it immediately becomes also a solution of \eqref{eq:eom(2a)}, up to higher orders in $\pi(x)$, given the pion's equation of motion \eqref{eq:pi_eom(1)}.

We can also expand the equations of motion \eqref{eq:eom(2)} in flat-space derivatives:
\begin{subequations}\label{eq:eom(2,1)}
\begin{align}
\d\cdot V^{(2,1)}&=0~, \label{eq:eom(2,1a)} \\
 z\d_z\frac{1}{z}\dz\Vmat{2,1}&=-iz^2\big[\pi,\dm\pi\big]~,\label{eq:eom(2,1b)}
\end{align}
\end{subequations}
where now the superscript $(m,n)$ identifies a term with $m$ pions and $n$ (flat-space) derivatives. With BCs (\ref{eq:BCs_UV}-\ref{eq:BCs_IR}), \eqref{eq:eom(2,1b)} yields
\begin{align}\label{eq:solution(2,1)}
& \Vmat{2,1}=i\vat{2,1}\big[\pi,\dm\pi\big]~,~~~\vat{2,1}=-\frac{1}{8}z^2(z^2-2)~,
\end{align}
which is also a solution to \eqref{eq:eom(2,1a)} under \eqref{eq:pi_eom(1)}.
Iterating the procedure we can expand \eqref{eq:eom(2)} to next-to-leading order in derivatives
\begin{subequations}\label{eq:eom(2,3)}
\begin{align}
 \d\cdot V^{(2,3)}&=0~,\label{eq:eom(2,3a)}\\
 \Dz\left(\dz\Vmat{2,3}\right)&=\dnu\left(\dn\Vmat{2,1}
 -\dm\Vmat{2,1}\right)~,\label{eq:eom(2,3b)}
 %=2i\vat{2,1}\Big[\dnu\pi,\dn\dm\pi\Big]~,
\end{align}
\end{subequations}
where again the r.h.s. is written in terms of the previous solution \eqref{eq:solution(2,1)}.  From \eqref{eq:eom(2,3)} we instantly get
\begin{align}\label{eq:solution(2,3)}
&\Vmat{2,3}= i\vat{2,3}\big[\dnu\pi,\dn\dm\pi\big]~,~~~\vat{2,3}=-\frac{1}{96}z^2(z^2-3)^2~,
\end{align}
which, together with \eqref{eq:solution(2,1)}, agrees with \eqref{eq:solution(2)}.

\textbf{Third order solution:} Expanding \eqref{eq:eom} \textit{naively} to $O(\pi^3)$ we find
\begin{subequations}\label{eq:eom(3)}
\begin{align}
 \dmu\big(\dm\Azat{3}-\dz\Amat{3}\big)&=-2i\dmu\big[\Azat{1},\Vmat{2}\big]~,\label{eq:eom(3a)}\\
 \Dz\big(\dm\Azat{3}-\dz\Amat{3}\big)+\dnu\big(\dn\Amat{3}-\dm\Anat{3}\big)&=
 -2iz\dz\frac{1}{z}\big[\Azat{1},\Vmat{2}\big]~.\label{eq:eom(3b)}
\end{align}
\end{subequations}
Note that $\Az$ and $\Am$ only enter the l.h.s. through the combination \makebox{$\dm\Az-\dz\Am$}. For $n\geq 1$, we can use the remaining gauge freedom to set $\Azat{2n+1}=0$ at each order, and shift the entire contribution to $\Amat{2n+1}$. This amounts, as stated earlier, to setting \eqref{eq:Az_gauge_fix}
as an exact gauge fixing.
Expanding \eqref{eq:eom(3)} in derivatives and solving  the $\Am$ equation first, we find at leading order
\begin{align}\label{eq:solution(3,1)}
\Amat{3,1}= \aat{3,1}(z)\Big[\pi,\big[\pi,\dm\pi\big]\Big]
~,~~~\aat{3,1}(z)=\frac{1}{24}z^2(z^2-1)(z^2-2)~,
\end{align}
and at next-to-leading order
\begin{align}\label{eq:solution(3,3)}
& \Amat{3,3}=\aat{3,3}^{(1)}(z)\Big[\dnu\pi,\big[\dn\pi,\dm\pi\big]\Big]+\aat{3,3}^{(2)}(z)\Big[\pi,\big[\dnu\pi,\dn\dm\pi\big]\Big]~,\nn\\
& \aat{3,3}^{(1)}=\frac{1}{384}z^2(z^2-1)(z^4-5z^2+7)~,~~~ \aat{3,3}^{(2)}=\frac{1}{192}z^2(z^2-1)(z^2-3)^2~.
\end{align}
Plugging the solution into the $\Az$ equation seems to lead to a contradiction! This is because we have forgotten a term in \eqref{eq:eom(3)}. Indeed, when expanded to first order in $\pi(x)$, we used \eqref{eq:pi_eom(1)} to neglect the term $\d^2\Az=z\d^2\pi=O(\pi^3)$; at this order in pions, the $O(\pi^3)$ piece becomes  relevant. The expansion in the presence of $\pi$'s equation of motion should be treated with care, since the equation itself \eqref{eq:pi_eom(1)} shuffles the various orders. Adding the forgotten term $\big(\d^2 \Azat{1}\big)^{(3)}$ into the l.h.s. of \eqref{eq:eom(3a)} we find that \eqref{eq:solution(3,1)} and \eqref{eq:solution(3,3)} are also solutions of \eqref{eq:eom(3a)}, provided a unique correction to $\pi$'s equation of motion at $O(\pi^3)$,
\begin{align}\label{eq:pi_eom(3,4)}
\d^2\pi=\frac{1}{6}\Big[\dmu\pi,\big[\pi,\dm\pi\big]\Big]+\frac{11}{192}\Big[\dnu\pi,\big[\dmu\pi,\dn\dm\pi\big]\Big]+O(\pi^5,\d^6\pi^3)~.
\end{align}

It is straightforward to iterate this procedure and obtain the solution up to any finite order in derivatives and in pions. 
At any even order in $\pi$ there is a unique solution to $\Vm$'s equations of motion (and BCs), which is then automatically a solution to $\Vz$'s equation of motion, when using $\pi$'s equation of motion at lower orders. At each odd order in $\pi$ there is a unique solution to the $\Am$ equation of motion. That solution also solves the $\Az$ equation of motion, provided a unique correction to the equation of motion of $\pi$. The role of the $\Az$ and $\Vz$ equations of motion is thus only to enforce the pion's equation of motion in the effective theory. The solution can also be obtained systematically to any finite order in $\pi$, and to all orders in derivatives at once, using Green's functions for the operators on the l.h.s. of the equations of motion. However, this method is slightly more technical and we skip it here. Later on we reformulate everything in terms of Feynman diagrams, which essentially does the same thing but in a simpler fashion.

\textbf{On-shell effective action:}
We now plug the solutions obtained above back into the bulk action, and explicitly perform the integration in $z$. This leave us with a four-dimensional action for $\pi(x)$. As explained before, this is identified with the low-energy on-shell effective action for the pions. 
 There is no quadratic term in the expansion of \eqref{eq:bulk_action_on-shell}, as expected for an on-shell action.  The term giving an $\cO(\pi^4)$  contribution is
\begin{align}\label{eq:bulk_action_on-shell(4)}
\cS_{\pi^4}^{(5d,o.s.)}&=\frac{i}{g_4^2}\int_0^1 \frac{dz}{z}\int d^4x\Tr\Big\{\big[\Vmat{2},\Azat{1}\big]\dmu\Azat{1}\Big\}~.
\end{align}
Plugging in \eqref{eq:solution(1)final}\eqref{eq:solution(2,1)}\eqref{eq:solution(2,3)} and integrating over $z$ we find
\begin{subequations}\label{eq:pi_action_on-shell(4,2-4)}
\begin{align}
\cS_{\d^2\pi^4}&=-\frac{L^2}{24g_4^2} \int d^4x\Tr\Big\{\big[\pi,\dmu\pi\big]\big[\pi,\dm\pi\big]\Big\}~,\label{eq:pi_action_on-shell(4,2)}\\ 
\cS_{\d^4\pi^4}&=-\frac{11L^4}{768g_4^2} \int d^4x\Tr\Big\{\big[\dm\pi,\dn\pi\big]\big[\dmu\pi,\dnu\pi\big]\Big\}~,\label{eq:pi_action_on-shell(4,4)}
\end{align}
\end{subequations}
where we have reinstated the IR cutoff scale $L$.\footnote{Notice that this scale controls the derivative expansion.} We can also insert the all-order-in-derivatives result for $\Vm$ \eqref{eq:solution(2)} into \eqref{eq:bulk_action_on-shell(4)}  to find
\begin{align}\label{eq:pi_action_on-shell(4)}
\cS_{\pi^4}&=-\frac{1}{g_4^2}\int d^4x \Tr \bigg\{\big[\pi,\dm\pi\big]\frac{1}{4\d^2}\bigg(1-\frac{8}{(L\d)^2}\frac{J_2(L\d)}{J_0(L\d)}\bigg)\big[\pi,\dmu\pi\big]\bigg\}~,
\end{align}
which agrees with \eqref{eq:pi_action_on-shell(4,2-4)}. This is the exact on-shell effective action which controls the dynamics of four external pions through all energies (at infinite $N$)! 
Similarly, we derive the six-pion, two-derivative term,
\begin{align}\label{eq:pi_action_on-shell(6,2)}
\cS_{\d^2\pi^6}&=-\frac{L^4}{360g_4^2} \int
d^4x\Tr\bigg\{\Big[\pi,\big[\pi,\dm\pi\big]\Big]\Big[\pi,\big[\pi,\dmu\pi\big]\Big]\bigg\}~.
\end{align}

\subsection{Comparing with Hirn and Sanz}\label{subsec:comparison}
We have derived above the on-shell effective action for the pions in the Hirn-Sanz model \cite{Hirn:2005nr}. Our results superficially differ from those of Hirn and Sanz \cite{Hirn:2005nr} in three ways.  Our effective action is written in terms of our holographic pion, $\pi$, whereas the authors of \cite{Hirn:2005nr} use the NG pion, $\Pi$, defined in \eqref{eq:true_pion}. That means the two effective actions should only agree on-shell.  Moreover, \cite{Hirn:2005nr}  explicitly express their result not directly in terms of $\Pi$, but rather in terms of $U=e^{i\Pi/f_\pi}$. Finally, we have an on-shell effective action, instead of the off-shell action of \cite{Hirn:2005nr}. 
 In Appendix \ref{app:off-shell} we also compare our results at the off-shell level, from which we derive the precise relation between the two pions. We find a very simple relation, which could have been anticipated as explained below.
 
As in \eqref{eq:overlap}, our pion mode $\pi(x)$, for which we compute the effective action, is related to the NG pion $\Pi(x)$ via
\begin{align}
\pi=c\Pi+O(\Pi^3)~,
\end{align}
with some yet unknown $c$.
 The pion of Hirn and Sanz \cite{Hirn:2005nr} is defined through its exponential,
\begin{align}\label{eq:true_pion}
U\equiv \exp{\big[i\Pi/f_{\pi}\big]}
\end{align}
($f_{\pi}$ is the pion decay constant), which transforms \textit{covariantly} under the spontaneously broken flavor symmetry,
\begin{align}\label{eq:U_trans_law}
U\rightarrow LUR^{\dagger}~.
\end{align}
To make the comparison, we first convert the chiral Lagrangian from the $U$ language to the $\pi$ language, and then evaluate it on-shell by deriving its equations of motion and plugging them back into the action.

The chiral effective Lagrangian is written at lower orders in terms of derivatives of $U$.\footnote{Both $\d$ and $\pi$ scale with $E/f_\pi$ for a process at energy scale $E$. This is the small parameter in the expansion.} The unique two-derivative term is
\begin{align}\label{eq:chiralkinetic}
\cL_2=\frac{f_{\pi}^2}{4}\dm \Ud \dmu U~,
\end{align}
 by which the pion $\Pi$ obtains a canonical kinetic term (given the trace conventions in Appendix \ref{app:conventions} and the definition of \eqref{eq:true_pion}).
 Expanding this term up to six-pion order and
 putting it on shell we find
 \begin{align}\label{eq:on-shell_chiral_ (2,4)}
\cL_{2,4}^{(o.s.)}+\cL_{2,6}^{(o.s.)}=\Tr\bigg\{-\frac{1}{48
f_{\pi}^2}\big[\Pi,\dm\Pi\big]\big[\Pi,\dmu\Pi\big] -\frac{1}{720
f_{\pi}^4}\Big[\Pi,\big[\Pi,\dm\Pi\big]\Big]\Big[\Pi,\big[\Pi,\dmu\Pi\big]\Big]\bigg\}~.
\end{align}
Matching to our result at the two-derivative order \eqref{eq:pi_action_on-shell(4,2-4)}, and remembering the overall normalization between the two modes, we find
\begin{align}
\frac{L_1^2c^4}{24g_4^2}=\frac{1}{48f_{\pi}^2}~,~~~\frac{L^4c^6}{360g_4^2}=\frac{1}{720f_{\pi}^4}~.
\end{align}
From this we find $f_{\pi}$ and $c$,
\begin{align}\label{eq:fpi}
f_{\pi}=\frac{\sqrt{2}}{g_4L}~,~~~c=\frac{g_4}{\sqrt{2}}=\frac{1}{f_{\pi}L}~,
\end{align}
in a perfect agreement with the pion decay constant of Hirn and
Sanz.
At the four-derivative order, the chiral Lagrangian for $N_f=3$ (a condition specifically used in \cite{Hirn:2005nr}) and with no background fields turned on consists of three independent terms \cite{Gasser:1983yg,Gasser:1984gg},
\begin{align}
\cL_4=L_1\big\langle \dm \Ud \dmu U \big\rangle^2+L_2\big\langle \dm \Ud \dn U
\big\rangle\big\langle \dmu \Ud \dnu U \big\rangle + L_3\big\langle \dm \Ud \dmu U
\dn \Ud \dnu U \big\rangle~,
\end{align}
where $\langle ... \rangle$ stands for the flavor trace. Expanding
in $\Pi$'s and evaluating on shell, at four-pion order we find
\begin{align}\label{eq:chL_on-shell(4,4)}
\cL_{4,4}^{(o.s.)}=-\frac{L_1}{f_{\pi}^4}\big\langle \dm \Pi \dmu \Pi
\big\rangle^2-\frac{L_2}{f_{\pi}^4}\big\langle \dm \Pi \dn \Pi
\big\rangle\big\langle \dmu \Pi \dnu \Pi \big\rangle -\frac{L_3}{f_{\pi}^4}
\big\langle \dm \Pi \dmu \Pi \dn \Pi \dnu \Pi \big\rangle~.
\end{align}
We can  now rewrite our result \eqref{eq:pi_action_on-shell(4,4)} in the form of \eqref{eq:chL_on-shell(4,4)}.
For flavor group $SU(3)$ we have the following identity (for example, see \cite{Donoghue:1992dd}),
\begin{align}\label{eq:algebra}
&\Big\langle \big[\dm\pi,\dn\pi\big]\big[\dmu\pi,\dnu\pi\big]\Big\rangle=2\big\langle \dm\pi\dn\pi\dmu\pi\dnu\pi\big\rangle-2\big\langle \dm\pi\dmu\pi\dn\pi\dnu\pi\big\rangle\nn\\
&=\big\langle\dm\pi\dmu\pi\big\rangle^2+2\big\langle\dm\pi\dn\pi\big\rangle\big\langle\dmu\pi\dnu\pi\big\rangle-6\big\langle \dm\pi\dmu\pi\dn\pi\dnu\pi\big\rangle~.
\end{align}
Using \eqref{eq:algebra} and \eqref{eq:fpi} in \eqref{eq:pi_action_on-shell(4,4)} our on-shell action reads
\begin{align}\label{eq:chiralL2}
\cS_{4d}^{(\d^4\pi^4)}=\frac{11}{768g_4^2f_{\pi}^4}\int d^4x \Tr\Big\{6\big\langle\d_{\mu}\pi\d^{\mu}\pi\d_{\nu}\pi\d^{\nu}\pi\big\rangle-\big\langle\d_{\mu}\pi\d^{\mu}\pi\big\rangle^2
-2\big\langle\d_{\mu}\pi\d_{\nu}\pi\big\rangle\big\langle\d^{\mu}\pi\d^{\nu}\pi\big\rangle\Big\}~.
\end{align}
Matched with \eqref{eq:chL_on-shell(4,4)}, we get
\begin{align}\label{eq:solution_Li's}
L_2=2L_1~,~~L_3=-6L_1~,~~L_1=\frac{11}{768g_4^2}~,
\end{align}
again in a perfect agreement with Hirn and Sanz. 

We should point out one implicit (and unimportant) difference compared to the Hirn and Sanz result. While we have defined $g_4^2\equiv g_5^2/R$, they make a slightly different definition $g_4^2\equiv g_5^2/l_0$, where $l_0$ is the regularization parameter cutting off $z$ near the UV of the  bulk geometry, with $l_0$ eventually sent to zero. This is only a redefinition of the arbitrary scale $f_{
\pi}$. In other words, we could have absorbed the difference in our $g_4$ into  $c$, to match the Hirn-Sanz result exactly.

In sum, we used our prescription to straightforwardly compute the chiral Lagrangian coefficients at order $\d^4$ in the Hirn-Sanz AdS/QCD model, and found perfect agreement with previous results.
It is now a simple matter to extend our results to higher orders and compare with those in the chiral Lagrangian expansion, in order to compute higher order coefficients. (In fact, \eqref{eq:pi_action_on-shell(4)} already contains an infinite number of independent chiral Lagrangian coefficients,  equivalent to the exact four-pion scattering amplitude for this model, which we obtain explicitly below.) 

\section{Diagrammatics}\label{sec:diagrammatics}

The procedure described above provides a
straightforward method for deriving the pion effective action: we define the pion to be the IR-boundary value of the radial component of the axial bulk gauge field; we  solve the classical equations of motion for all bulk fields  order by order in the pion's magnitude; finally, we plug these solutions back into the action and integrate over the radial direction. We can also derive the same on-shell effective action (or the S-matrix) directly using tree-level Feynman diagrams with pions on the external legs. Since the pions are explicitly defined as the boundary values of $A_z$, pictorially we will have ``inverted Witten diagrams", i.e.  bulk Feynman diagrams that start and end on the IR boundary.\footnote{Note that our gauge fixing $\Az(x,z)\rightarrow z\pi(x)$ will qualitatively change this picture. After fixing this gauge, the pion will have non-zero wavefunction along the whole radial direction.} This method proves more economical than the pedestrian approach of the previous section, as it obviates the need to solve for $A_\mu,V_\mu$, and instead directly obtains $S_{eff}^{(on-shell)}$. 

Consider again the bulk partition function,
\begin{align}\label{eq:mode_separation}
Z_{bulk}^{}=\int D\pi \int_{A_z(x,L)=\pi(x)} DA_z DA_{\mu}
DV_{\mu}e^{iS[A_z,V_\mu,A_\mu]}  ~.
\end{align}
Our prescription identifies the pion effective action as
\begin{align}\label{eq:pi_effective_action}
e^{iS_{eff}^{(on-shell)}[\pi]}=\int_{A_z(x,L)=\pi(x)} DA_z DA_\mu
DV_\mu e^{iS[A_z,V_\mu,A_\mu]}  ~.
\end{align}
This is the analog of \eqref{eq:prescription2} for the AdS/QCD example of section \ref{sec:example}. The bulk fields are also subject to the boundary conditions (\ref{eq:BCs_UV}), (\ref{eq:BCs_IR}).
We have already imposed the gauge condition $V_z=0$ in writing (\ref{eq:mode_separation}). The exclusive identification of $A_z$'s boundary value with the pion mode is also not gauge-invariant, and we consider the particular splitting we
use to be part of the gauge choice. (Note also that because we work strictly at tree level, we can safely neglect ghosts.) Nevertheless, some gauge freedom remains to be fixed. Given the gauge-symmetry-breaking boundary condition for $\Am$ \eqref{eq:BCs_IR} and the boundary-value assignment for $\Az$  \eqref{eq:our_assignment}, a rigorous gauge fixing procedure is somewhat tricky. However, under the classical equations of motion we have shown that $\Az=z\pi$ constitutes a legal gauge choice. Since we eventually compute on-shell observables (such as scattering amplitudes) at tree level, we can try using this gauge fixing inside the path integral, and hope it gives the correct result. For now, we take this as our ansatz. We will see later that this works modulo a subtlety which arises at $O(\pi^6)$; we return to this point in more detail below. A simplified form of the partition function is then
\begin{align}\label{eq:pi_off_shell_effective_action}
e^{iS_{eff}^{(off-shell)}[\pi]}= \int DA_{\mu}
DV_{\mu} e^{iS[A_z = z\pi(x),V_\mu,A_\mu]}  ~.
\end{align}
Since we apply the gauge fixing $A_z=z\pi$ and ignore the path integration over $A_z$, we are now  computing an \textit{off-shell} effective action for $\pi$. We can obtain the on-shell action either by deriving the resulting EOM and plugging them back in, as shown in the previous section,  or by also including tree-level diagrams with propagating pions, governed by the  off-shell effective action we obtain (which includes a kinetic term). Alternatively, by taking the pions to be in asymptotic states, we can directly compute their S-matrix.

 First, we find the kinetic term in the off-shell effective action by evaluating $A_z$'s kinetic term subject to the gauge fixing,
\begin{align}
S_{\pi^2}^{(off-shell)}[\pi]=\frac{1}{2g_4^2}\int d^4x
\d_\mu\pi^a\d^\mu\pi^a~.
\end{align}
We can obtain the $n$-pion interaction term, meanwhile, by summing all connected tree-level diagrams with $V_\mu$'s and $A_\mu$'s propagating on internal legs, and with $n$ external  $\pi$ legs (see for instance figure \ref{fig:4pi}). The 5d Feynman rules and Green's functions are straightforward to compute, and are summarized in Appendix \ref{app:Feynman}. Below, we show explicitly how
to obtain the four-pion and six-pion effective vertices.
We compare the four-pion term to
our results from the previous section, and derive the six-pion term in the effective action to all orders in derivatives, a novel result.

\subsection{Four-Pion Effective Action}
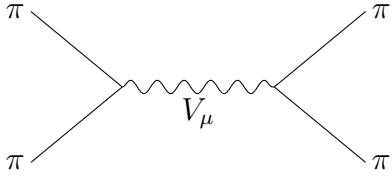
\begin{figure}\label{fig:4pi}
\begin{center}
\begin{tikzpicture}
\path (-1,0) node (Va) {}
    (1,0) node (Vb) {}
    (-2.2,1.) node (aup) {$\pi\quad$}
    (-2.2,-1) node (adown)  {$\pi\quad$}
    (2.2,1) node (bup)  {$\quad\pi$}
    (2.2,-1) node (bdown)  {$\quad\pi$};
\draw[-,decorate,decoration=snake] (-1.,0) -- (1.,0) node[below,midway] {$V_\mu$}  ;
\draw[-]  (-1,0) to (-2.2,1)  ;
\draw[-] (-1,0) to (-2.2,-1) ;
\draw[-]  (1,0) to (2.2,1) ;
\draw[-] (1,0) to (2.2,-1);
\end{tikzpicture}
\caption{The only 5d diagram contributing to the effective four-pion vertex. Modulo the mentioned gauge-fixing subtlety, the above process depicts external pions that live on the IR boundary, which interact with a gauge field that lives in the bulk. }
\end{center}
\end{figure}

There is only one diagram that contributes to the four-pion term, shown in Figure \ref{fig:4pi}.
 Using the Feynman rules
described in Appendix \ref{app:Feynman} we  find\footnote{Since we work at most to sixth order in pions, we are allowed to put the external pions on-shell at leading order in $\pi$. In other words,
we can set $k_1^2=k_2^2=0$, which in turn implies that the longitudinal piece of the vector propagator does not contribute at this order. Such terms may contribute at order $\pi^8$ and higher, and should be taken into account in those cases.}
\begin{align}
S_{\pi^4}^{(off-shell)}
&=-\frac{1}{24g_4^2L^4}\int\frac{d^4k_1 d^4k_2 d^4k_3 d^4k_4
}{(2\pi)^{16}}\delta^{(4)}(k_1+k_2+k_3+k_4) \prod\limits_{i=1}^4 \pi^{a_i}(k_i)\nonumber\\
& \hspace{2cm}\phantom{+}\Big[ f^{a_1a_2e}f^{a_3a_4e} I_4(k_{12})(k_1-k_2)\cdot(k_3-k_4)\nonumber\\
 &\hspace{2cm}+ f^{a_1a_3e}f^{a_2a_4e} I_4(k_{13})(k_1-k_3)\cdot(k_2-k_4)\nonumber\\
 &\hspace{2cm}+  f^{a_1a_4e}f^{a_2a_3e} I_4(k_{14})(k_1-k_4)\cdot(k_2-k_3)\Big] \label{eq:4ptaction}\\
&=-\frac{1}{g_4^2L^2}\int d^4x \Tr \big[\pi,\d^\mu\pi\big]I_4(i\d)\big[\pi,\d_\mu\pi\big]~,
\end{align}
where
$\d\equiv\sqrt{\d^2}$ and
we define
\begin{align}
k_{ij}\equiv\sqrt{(k_i+k_j)^2}~.
\end{align}
The vector Green's function is defined in equation (\ref{eq:Vprop}), and
the integral  $I_4$ is
\begin{align}
I_4(k)\equiv\int_0^1 dz\int_0^1 dz' zz' G_V(k,z,z') &= \frac{1}{4k^2}\left[ 1-\frac{8J_2(k)}{k^2J_0(k)}\right]\\
&= -\frac{1}{24} -\frac{11}{1536}k^2 -\frac{19}{15360}k^4+\cO(k^6)~.
\end{align}
The function $I_4$ has poles at the zeroes of $J_0$, coinciding (as they should) with
the masses of vector meson states in the Hirn-Sanz model.
We thus have the off-shell action up to fourth order in pions and
to all orders in derivatives,
\begin{align}\label{eq:pieff4}
\cS_{\pi^4}^{(off-shell)} &= \frac{1}{2g_4^2}\int d^4x  \ \Big\{ \Tr \d_\mu\pi\d^\mu\pi -2 \Tr \big[\pi,\d^\mu\pi\big]I_4(i\d)\big[\pi,\d_\mu\pi\big]\Big\}
\end{align}
It is straightforward to verify that this action  produces the same on-shell action derived in equation \eqref{eq:pi_action_on-shell(4)},  by writing the pion's equation of motion
\begin{align}
\d^2\pi =  -4\Big[\d_\mu\pi, I_4(i\d)\big[\pi,\d^\mu\pi\big]\Big]~,
\end{align}
 and plugging it back into the action. By using the Feynman rules derived from (\ref{eq:pieff4}) we also find the four-pion scattering amplitude to all orders in derivatives:
 \begin{align}
\cM_{\pi\pi\rightarrow\pi\pi}
=  \frac{g_4^2}{L^4}
\Big[f^{a_1a_2c} f^{a_3a_4c}I_4(\sqrt{s})(u-t) 
+f^{a_1a_3c} f^{a_2a_4c}I_4(\sqrt{t})(s-u)
+f^{a_1a_4c} f^{a_2a_3c}I_4(\sqrt{u})(t-s)    \Big]~,
\end{align}
in terms of the standard Mandelstam variables: $s=k_{12}^2~,t=k_{13}^2~,u=k_{14}^2~.$

\subsection{Six-Pion Effective Action}
\begin{figure}\label{fig:6pi}
\begin{center}
\begin{tikzpicture}
\path (0,-2.5) node (label) {(1)}
    (0,0) node (aV0) {}
    (-1.299,-.75) node (aV1) {}
    (1.299,-.75) node (aV2) {}
    (0,1.5) node (aV3) {};
\draw[-,decorate, decoration=snake] (0,0) -- node {$ \qquad V_\mu$}  (0,1.5)  ;
\draw[-,decorate, decoration=snake] (0,0) -- node[below] {$V_\mu$}  (-1.299,-.75);  
\draw[-,decorate, decoration=snake] (0,0) -- node[below] {$V_\mu$} (1.299,-.75);
\draw[] (-1.299,-.75) to  (-1.558,-1.716) node[below] {$\pi$} ;
\draw[] (-1.299,-.75) to    (-2.265,-0.491)  node[left] {$\pi$}  ;
\draw[] (1.299,-.75) to  (2.265,-.491) node[right] {$\pi$} ;
\draw[] (1.299,-.75) to    (1.558,-1.716)  node[below] {$\pi$}  ;
\draw[] (0,1.5) to  (.707,2.207)  node[right,above] {$\quad\pi$}  ;
\draw[] (0,1.5) to    (-.707,2.207) node[left,above] {$\pi\quad$}  ;
\end{tikzpicture}\hspace{1cm}
\begin{tikzpicture}
\path  (0,-1.5) node (label) {(2)}
(-.75,0) node (bV1){}
 (.75,0) node (bV2){}
 (-1.5,1.299) node (bV3){}
 (1.5,1.299) node (bV4){};
\draw[-,decorate,decoration=snake] (-.75,0) to (.75,0) node[above, midway] {$A_\mu$};
\draw[-,decorate,decoration=snake] (-.75,0) -- node[left] {$V_\mu$} (-1.5,1.299) ;
\draw[-,decorate,decoration=snake] (.75,0) -- node[right] {$\ V_\mu$} (1.5,1.299) ;
\draw[] (1.5,1.299) to (2.466,1.558) node[right] {$\pi$};
\draw[] (1.5,1.299) to (1.241,2.265) node[left,above] {$\pi$}  ;
\draw[] (-1.5,1.299) to (-2.466,1.558) node[right,above] {$\pi\quad$};
\draw[] (-1.5,1.299) to (-1.241,2.265) node[above] {$\pi$}  ;
\draw[] (-.75,0) to (-1.25,-.866) node[below] {$\pi$}  ;
\draw[] (.75,0) to (1.25,-.866) node[below] {$\pi$}  ;
\end{tikzpicture}\hspace{1cm}
\begin{tikzpicture}
\path  (0,-2.5) node (label) {(3)}
    (0,0) node (cV1) {}
    (.707,.707) node (cV2) {}
    (-.707,.707)  node (cV3) {}
       (-1.061,-1.061)  node (cV3) {}
      (1.061,-1.061)  node (cV4) {};
\draw[-,decorate, decoration=snake] (0,0) -- node[left] {$V_\mu \ $}  (-1.061,-1.061);
\draw[-,decorate, decoration=snake] (0,0) -- node[right] {$ \ V_\mu$} (1.061,-1.061);
\draw[] (0,0) to (.707,.707) node[above] {$\pi$};
\draw[] (0,0) to (-.707,.707) node[above] {$\pi$};
\draw[] (1.061,-1.061) to (1.061,-2.061) node[below] {$\pi$};
\draw[] (1.061,-1.061) to (2.061,-1.061) node[right] {$\pi$};
\draw[] (-1.061,-1.061) to (-1.061,-2.061) node[below] {$\pi$};
\draw[] (-1.061,-1.061) to (-2.061,-1.061) node[left] {$\pi$};
\end{tikzpicture}
\caption{The only three diagrams from the 5d action that contribute to the six-pion interaction term (up to exchanges of external pion legs). Diagrams (2) and (3) contribute at $O(\d^2)$ in the derivative expansion, while Diagram (1) only begins to contribute at $O(\d^4)$.}
\end{center}
\end{figure}
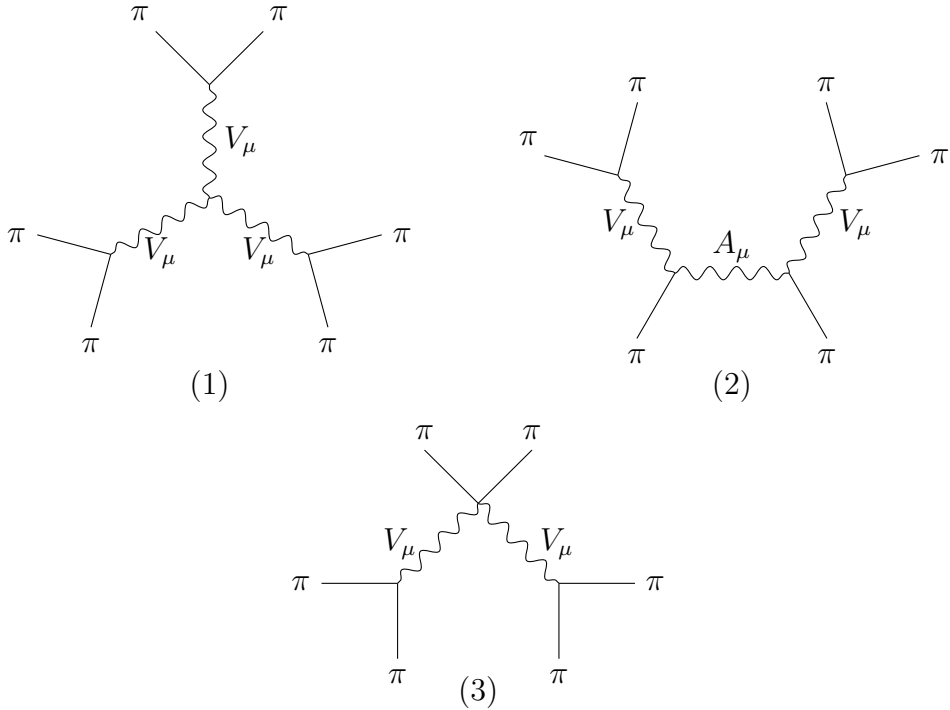

The power of this method becomes evident when computing higher order terms. Here we outline the computation of the six-pion term, which
we can easily compute to arbitrary order in derivatives, though we are unable to write it in a closed analytic form. The six-pion term in the effective action is obtained from
tree-level diagrams with six external pions. There are only three such diagrams (up to relabeling of external pion legs). These diagrams are shown in Figure \ref{fig:6pi}.

Each diagram is characterized by some 4d Lorentz structure and some $z$-integrals. The integrals corresponding to the diagrams in Figure \ref{fig:6pi} are given by
\begin{align}\label{eq:6piints}
I_{6,1}(k_a,k_b,k_c) &= \int\frac{dz_d}{z_d} dz_a z_a   dz_b z_b dz_c z_cG_V(k_a,z_a,z_d)G_V(k_b,z_b,z_d)    G_V(k_c,z_c,z_d)~, \\
I_{6,2}(k_a,k_b,q)&=\int dz_az_adz_bz_bdz_cdz_d\d_{z_c}G_V(k_a,z_a,z_c)\d_{z_d}G_V(k_b,z_b,z_d)G_A(q,z_c,z_d)~, \label{eq:I_{6,2}}\\
I_{6,3}(k_a,k_b) &= \int  dz_a z_adz_b z_bdz_c z_c G_V(k_a, z_a,z_c) G_V(k_b, z_b,z_c)~.
\end{align}
 (The details are relegated to Appendix \ref{app:6pi}.)
Summing the  contributions from all three diagrams, we find,
\begin{align}\label{eq:fullS6}
S_{\pi^6}^{(off-shell)}=&-\frac{R}{g_5^2}\int \prod\limits_{i=1}^6dk_i \ \Tr\bigg\{ \Big[\pi(k_5),\big[\pi(k_1),\pi(k_2)\big]\Big] \Big[\pi(k_6),\big[\pi(k_3),\pi(k_4)\big]\Big]\bigg\}\times\nonumber\\
&\qquad\Bigg\{ (k_1-k_2)\cdot(k_3-k_4)\bigg[ \frac{1}{4}I_{6,3}(k_{12},k_{34})+I_{6,2}(k_{12},k_{34},|k_1+k_2+k_5|)\bigg]\nonumber\\
&\qquad~~- \big(k_5\cdot (k_1-k_2)\big)\big (k_6\cdot (k_3-k_4)\big) \frac{I_{6,2}(k_{12},k_{34},|k_1+k_2+k_5|)-I_{6,2}(k_{12},k_{34},0)}{(k_1+k_2+k_5)^2}\nonumber\\
&\qquad~~+\frac{1}{6} I_{6,1}(k_{12},k_{34},k_{56})\Big[ (k_1+k_2)\cdot (k_5-k_6)(k_1-k_2)\cdot (k_3-k_4)\nonumber\\
&\qquad\qquad\qquad\qquad\qquad\qquad+(k_3+k_4)\cdot (k_1-k_2)(k_3-k_4)\cdot (k_5-k_6)\nonumber\\
&\qquad\qquad\qquad\qquad\qquad\qquad+(k_5+k_6)\cdot (k_3-k_4)(k_5-k_6)\cdot (k_1-k_2)\Big]\Bigg\}
\end{align}
This result is exact to all orders in derivatives.
To compare with our findings from section \ref{sec:example}, we can expand order by order in derivatives, and rewrite the interaction term in position space. The leading term is
\begin{align}
S_{\d^2\pi^6}^{(off-shell)} = \frac{L^4}{720g_4^2} \int
d^4x\Tr\bigg\{\Big[\pi,\big[\pi,\dm\pi\big]\Big]\Big[\pi,\big[\pi,\dmu\pi\big]\Big]\bigg\}~,
\end{align}
in full agreement with \eqref{eq:pi_action_on-shell(6,2)}, on shell. 

We should now discuss the subtlety mentioned earlier.  Diagram (2) in Figure \ref{fig:6pi} involves a vertex
coupling of the form $\int d^4x dz/z~\Az \Am \dz \Vmu$ which turns into $\int d^4xdz~\pi \Am\dz\Vmu$ after our gauge fixing. Naively, one could integrate the vertex by parts to arrive at $-\int d^4xdz~\pi\dz\Am\Vmu$,
which should then give the same final result. (Remember that $\pi(x)$ is strictly four-dimensional.) This does not turn out to be the case, however. The correct result is only obtained with the first form of the vertex above. The subtlety is due to the fact that $A_z=z\pi$ is not actually a valid  gauge-fixing inside the path integral. If we first integrate by parts in the action, and then use the same (strictly illegal) gauge-fixing procedure, we get different (and wrong) results.\footnote{Equivalently, in \eqref{eq:I_{6,2}} one cannot integrate by parts with respect to $z_c,z_d$ (as can be easily verified), since $G_A$ is not a smooth function. }   In other words, the  gauge-fixing ansatz we have considered is not fully consistent, and in particular does not commute with integration by parts. The exact form of the action needs to be unambiguously `chosen' to obtain the correct results. We leave the execution of an honest gauge fixing and the full exploration of this subtlety to future work.

\section{Summary and Outlook}\label{sec:discussion}
We have described a holographic method for deriving on-shell effective actions for large $N$ confining gauge theories. The effective action we compute is the one obtained by integrating out all massive degrees of freedom in the theory,  leaving only the massless  modes (including those with high energy). We identify the IR value of the appropriate bulk field with the massless mode of the gauge theory. Then, 
in the spirit of holographic Wilsonian renormalization,
 we integrate out all fields except for the fixed IR mode. Though the IR mode whose effective action we compute may not be exactly the same as the massless mode of the gauge theory, the universality of the S-matrix guarantees that their on-shell actions are the same. The on-shell effective action does not depend on the precise mixing between the bulk IR mode and the massless resonance, and is  thus independent of the  IR mode choice.  Although inspired by it, our results stand independently of the holographic Wilsonian RG formalism.

We demonstrated the mechanics of the procedure using the well-studied case of the Hirn-Sanz AdS/QCD model, where we can in principle generate S-matrix elements for
arbitrary numbers of pions and arbitrary orders in the momentum expansion. Concretely, we derived exact results for the four- and six-pion scattering amplitudes in this model, to all orders in momentum.

We have developed the technique for a very simple case: truncated AdS, with easily-identified Nambu-Goldstone modes. Though we have focused on massless states, the formalism can be used to compute effective actions for massive modes as well. It would be interesting to examine such situations, in which the bulk IR mode mixes with an infinite tower of resonances. 
The success of the
method should also not depend on the truncation of
 spacetime. We hope to extend our techniques to spacetimes in which a smooth gravitational potential induces confining behavior, such as the soft wall model
of AdS/QCD \cite{Karch:2006pv}, or other confining geometries, like the warped deformed conifold of \cite{Klebanov:2000hb}.

Our methods would find fruitful application in a variety of areas, from more complicated versions of holographic QCD or holographic technicolor, to duals of condensed matter systems. 
For example, this framework renders the study of additional (e.g. $(F_{\mu\nu})^n$) interaction terms quite simple, allowing one to concretely estimate the error introduced by neglecting such terms in
AdS/QCD frameworks.
One might also use such techniques to derive the low-energy effective action of fluctuations around holographic realizations of spatially inhomogeneous vacua (or ``striped phases'')
identified in AdS/QCD and AdS/CMT \cite{Ooguri:2010kt, Donos:2011bh}, especially in the confining phase \cite{Bayona:2011ab}. Another interesting application would be to follow our procedure in a black-brane geometry, in order to learn about the effective field theory of hydrodynamics \cite{Dubovsky:2011sj} holographically \cite{Amado:in_progress}. One can also use the same formalism to compute the effective action of other massless modes, for instance the effective action for the Goldstino in a holographic SUSY breaking scenario. 

Even more interestingly, it might be possible to use this method to engineer a holographic version of Seiberg duality. Several years ago \cite{Komargodski:2010mc} suggested relating the hidden local (flavor) symmetry (HLS) in SQCD with its ``emergent" Seiberg-dual gauge group. In particular, the vector rho meson is thus identified with the magnetic gauge  boson.\footnote{See \cite{Harada:1999zj} for a related early observation, and \cite{Kitano:2011zk,Abel:2012un} for further recent support.} In a generic, non-SUSY scenario -- e.g. QCD -- this gauge boson is Higgsed (together with all other vector mesons) and becomes the ordinary, massive rho meson. In the holographic context, vector mesons are realized as KK modes of the bulk gauge field dual to the flavor current. Indeed, the bulk gauge group, KK decomposed, can be associated with an infinite tower of hidden local symmetries \cite{Son:2003et}. Thus, finding a background in which the lowest KK mode of a bulk gauge field is actually massless would constitute a holographic realization of Seiberg duality. The magnetic theory would simply be the IR limit of the bulk theory, which is exactly dual to the electric theory. For related ideas see \cite{Abel:2010vb,Abel:2012un}. In such a case, we could use our methods to easily compute the effective action of the magnetic gauge boson, the Seiberg-dual action. 

Finally, it would be very interesting to use similar methods to learn more about holographic Wilsonian renormalization in general, and in particular about the precise field theory renormalization scheme that corresponds to a radial cutoff in the bulk. For example, starting with our procedure, one could ``un-integrate'' a thin bulk slice at the IR, trying to identify the gauge theory modes (for example, in terms of mesons) that are brought back to life.

\section*{Acknowledgments}
We are indebted to Zohar Komargodski for his collaboration in the early stages of this work,
for useful discussions and for his comments on the manuscript. We are very grateful to Ofer Aharony for many helpful discussions and detailed suggestions, and for his careful reading and useful comments on the manuscript. MF would also like to thank Rajesh Gopakumar for helpful discussions, and Harish-Chandra Research Institute for support during this work. Both MF and SKD were supported by the Weizmann Institute throughout much of this project.  
SKD is currently supported by the NYU Postdoctoral and Transition Program for Academic Diversity Fellowship. MF is supported in part by the Israeli Science Foundation under grant No. 392/09.

\appendix

\section{Conventions} \label{app:conventions} 
In order to facilitate comparison to \cite{Hirn:2005nr} we adopt their conventions,
\begin{align}
L_M = L_M^a\frac{T^a}{\sqrt{2}}\qquad\text{and}\qquad \Tr(T^aT^b)=2\delta^{ab}~.
\end{align}
The action 
\begin{align}
\cS_{5d}=-\frac{1}{4g_5^2}\int d^5x\sqrt{g} \ \Tr\Big( L_{MN}
L^{MN}+R_{MN}R^{MN} \Big)~
\end{align}
thus gives canonically normalized fields in 5d.
We work with the vector and axial-vector combinations,
\begin{align}
V_M = \half \left(L_M  + R_M\right)\qquad\text{and}\qquad A_M = \half\left(L_M - R_M\right)~,
\end{align}
and maintain the same normalization convention $A_M = \frac{1}{\sqrt{2}}A_M^a T^a~,$
and similarly for the pion, $\pi = \frac{1}{\sqrt{2}}\pi^aT^a$.

\section{Equations of Motion}\label{app:equationsofmotion}
The equations of motion in component form in the model of Hirn and Sanz, under the gauge condition $V_z=0$, are:
\begin{subequations}\label{eq:eom}
\begin{align}
&A_z:\quad
\dmu\Big(\dm\Az-\dz\Am\Big)=i\dmu\Big[\Vm,\Az\Big]+i\Big[\Vmu,\Amz\Big]
 +i\Big[\Amu,\Vmz\Big]~, \label{eq:eom_Az}\\
&A_\mu:\quad
z\d_z\frac{1}{z}\Big(\dm\Az-\dz\Am\Big)+\dnu\Big(\dn\Am-\dm\An\Big)=
 iz\dz\frac{1}{z}\Big[\Vm,\Az\Big]+i\Big[\Az,\Vmz\Big]\nn\\
 &\qquad\qquad+i\dnu\Big([\An,\Vm]+[\Vn,\Am]\Big)+i\Big[\Anu,\Vnm\Big]+i\Big[\Vnu,\Anm\Big]~,\label{eq:eom_Am}\\
&V_z:
\quad\dmu\Big(\dz\Vm\Big)=i\dmu\Big[\Az,\Am\Big]+i\Big[\Amu,\Azm\Big]
 +i\Big[\Vmu,\Vzm\Big]~, \label{eq:eom_Vz}\\
 &V_\mu: \quad z\d_z\frac{1}{z}\dz\Vm-\dnu\Big(\dn\Vm-\dm\Vn\Big)=
 iz\dz\frac{1}{z}\Big[\Az,\Am\Big]+i\Big[\Az,\Azm\Big]\nn\\
 &\qquad\qquad-i\dnu\Big([\An,\Am]+[\Vn,\Vm]\Big)
  -i\Big[\Anu,\Anm\Big]-i\Big[\Vnu,\Vnm\Big]~. \label{eq:eom_Vm}
\end{align}
\end{subequations}
As described in subsection \ref{subsec:prescription}, we solve these equations order by order in pion fields.

\section{All-Order Solution}\label{app:all-order solution}
Expanding \eqref{eq:eom_Vz},\eqref{eq:eom_Vm} at $O(\pi^2)$ and using the $O(\pi)$ solution, we have
\begin{align}\label{eq:expandedEOM2}
 \dmu\left(\dz\Vmat{2}\right)&=0~,\nn\\
  z\d_z\frac{1}{z}\dz\Vmat{2}-\dnu\left(\dn\Vmat{2}-\dm\Vnat{2}\right)&=
-i\Big[\Azat{1},\dm\Azat{1}\Big]~.
\end{align}
More generally, at this order
\begin{align}
\d\cdot V^{(2)}=0~.
\end{align}
The full equation at $O(\pi^2)$ is
\begin{align}\label{eq:expandedEOM2}
  &z\d_z\frac{1}{z}\dz\Vmat{2}-\d^2\Vmat{2}= -iz^2\big[\pi,\dm\pi\big]~.
\end{align}
We begin at $O(\pi^2)$ with the $\Vmat{2}$ equation \eqref{eq:expandedEOM2} and first solve for the homogenous part,
 \begin{align}
 \Dz\dz\bar\Vm-\d^2\bar\Vm=0~.
 \end{align}
Using separation of variables, we find the solution in terms of Bessel functions,
\begin{align}
\Vm (k,z)=z\big(c_1 J_1(kz)+c_2 Y_1(kz)\big)e^{ik\cdot x}\epsilon_\mu(k)~.
\end{align}
 Since we work in the absence of external vector sources, we keep only the normalizable mode $J_1$. Fourier-transforming we find the net homogeneous solution
\begin{align}
\bar \Vm(x,z)=\int\frac{d^4 k}{(2\pi)^4}zJ_1(kz)\epsilon_{\mu}(k)e^{ik\cd x}~,
\end{align}
where for reality of $\bar\Vm$ we have $\epsilon_{\mu}(-k)=\epsilon_{\mu}^*(k)$.

We now need to find a particular solution for the original equation.
We again solve by separation of variables,
$\hat\Vm=\xi(z)w_{\mu}(x)$~,
\begin{align}
\xi(z)\d^2 w_{\mu}(x)-\Dz\d_z\xi(z) w_{\mu}(x)= i z^2 \big[\pi(x),\dm\pi(x)\big]~.
\end{align}
Taking $\xi(z)=z^2$ we find
\begin{align}
&w_{\mu}(x)=\frac{i}{\d^2}\big[\pi,\dm\pi\big]~,
\end{align}
or, in Fourier space,
\begin{align}
w_{\mu}(x)=\int \frac{d^4k}{(2\pi)^4}\frac{e^{ik\cd x}}{k^2}\int\frac{d^4k'}{(2\pi)^4}
k'_{\mu}\big[\pi(k-k'),\pi(k')\big]~.
\end{align}
The most general solution to equation (\ref{eq:expandedEOM2}) is then $\Vmat{2}=\bar \Vm+\hat \Vm $,
\begin{align}
\Vmat{2}(x,z)=\int\frac{d^4 k}{(2\pi)^4}e^{ik\cd x}\bigg(zJ_1(kz)\epsilon_{\mu}(k)+ \frac{z^2}{k^2}\int\frac{d^4k'}{(2\pi)^4}k'_{\mu}\big[\pi(k-k'),\pi(k')\big]\bigg)~.
\end{align}
Imposing the boundary condition at  $z=1$ we have
\begin{align}
\dz \Vmat{2}(x,1)=\int\frac{d^4 k}{(2\pi)^4}e^{ik\cd x}\bigg(k J_0(k)\epsilon_{\mu}(k)+ \frac{2}{k^2}\int\frac{d^4k'}{(2\pi)^4}k'_{\mu}\big[\pi(k-k'),\pi(k')\big]\bigg)=0~.
\end{align}
(The boundary condition at $z=0$ is already satisfied.)
This uniquely sets the solution
\begin{align}
\epsilon_{\mu}(k)=\frac{-2}{k^3 J_0(k)}\int\frac{d^4k'}{(2\pi)^4}k'_{\mu}\big[\pi(k-k'),\pi(k')\big]~,
\end{align}
where we have used the identity
\begin{align}
\dz\big(z J_1(kz)\big)=kzJ_0(kz)~.
\end{align}
The full solution is then,
\begin{align}\label{eq:Vm2}
\Vmat{2}(x,z)=z^2\int\frac{d^4 k}{(2\pi)^4}\int\frac{d^4k'}{(2\pi)^4}k'_{\mu}\big[\pi(k-k'),\pi(k')\big]
\frac{1}{k^2}\bigg(1-\frac{2}{k z}\frac{J_1(kz)}{J_0(k)}\bigg)e^{ik\cd x}~.
\end{align}
Expanding the Bessel functions in $k$ we have,
\begin{align}
\frac{1}{k^2}\bigg(1-\frac{2}{k z}\frac{J_1(kz)}{J_0(k)}\bigg)=\frac{\left(z^2-2\right)}{8}
-\frac{\left(z^2-3\right)^2}{192} k^2+\frac{\left(z^6-12z^4+54z^2-76\right)}{9216}k^4+O(k^6)~.
\end{align}
Plugging it back into \eqref{eq:Vm2}, we get
 \begin{align}
\Vmat{2}(x,z)&=\frac{z^2}{8}\int\frac{d^4 k}{(2\pi)^4}\int\frac{d^4k'}{(2\pi)^4}k'_{\mu}\big[\pi(k-k'),\pi(k')\big]
\bigg\{\left(z^2-2\right)-\frac{1}{24}\left(z^2-3\right)^2k^2+O(k^4)\bigg\}e^{ik\cd x}\nn\\
&=-i\frac{z^2}{8}\bigg\{\left(z^2-2\right)-\frac{1}{24}\left(z^2-3\right)^2\d^2+O(\d^4)\bigg\}\big[\pi,\dm\pi\big]~,
\end{align}
which coincides with previous results.
Note that for this simple case we were easily able to guess the particular solution, while in general, one can solve the equations of motion in terms of
the Green's functions developed in Appendix \ref{app:GreenVmu}.

This result allows us to compute the four-pion on-shell action to all derivative orders directly by plugging \eqref{eq:Vm2} into \eqref{eq:bulk_action_on-shell(4)} and performing the $z$-integration without expanding the Bessel functions,
\begin{align}
\int_0^1 dz \frac{z^3}{k^2}\bigg(1-\frac{2}{k z}\frac{J_1(kz)}{J_0(k)}\bigg)=\frac{1}{4k^2}\bigg(1-\frac{8}{k^2}\frac{J_2(k)}{J_0(k)}\bigg)~.
\end{align}
The formal solution is
\begin{align}
\cS_{\pi^4}&=-\frac{R}{g_5^2}\int d^4x \Tr\bigg\{\big[\pi,\dm\pi\big]\frac{1}{4\d^2}\bigg(1-\frac{8}{\d^2}\frac{J_2(\d)}{J_0(\d)}\bigg)\big[\pi,\dmu\pi\big]\bigg\}\nn\\
&=\frac{R}{g_5^2}\int d^4x \Tr\bigg\{\big[\pi,\dm\pi\big]\bigg(-\frac{1}{24}+\frac{11}{1536}\d^2-\frac{19}{15360}\d^4+O(\d^6)\bigg)\big[\pi,\dmu\pi\big]\bigg\}~,
\end{align}
which coincides with previous results, and the result using Feynman diagrams described in section \ref{sec:diagrammatics}.
 Note that when using the equations of motion for $\pi$ this term will also contribute at higher orders in $\pi$, starting with $\d^4\pi^6$.

\section{Two-Derivative Terms to All Orders in Pions}

In the chiral Lagrangian, the spontaneously broken flavor symmetry is realized nonlinearly on the pions. That means, at fixed order in pions, a term in the action is not invariant under the full symmetry, but it can be made so under a unique completion to all orders in $\pi$ (separately for any derivative order). For example, whereas the kinetic term for pions in terms of $\pi$ is not invariant under the full symmetry, it can be made so by writing a kinetic term for 
$U\equiv\exp{\left(i\pi/f_{\pi}\right)}$. 

For a consistency check of our procedure, we solve for the two-derivative term in the action, to all orders in the pions, and verify that it is indeed consistent with the 
symmetry. 
To leading order in derivatives,  the equations of motion can be recast as
\begin{align}\label{eq:Amz_Vmz_EOM}
& \dz\left(\frac{1}{z}\Amz\right)=iz\left[\pi,\left(\frac{1}{z}\Vmz\right)\right]~,\nn\\
& \dz\left(\frac{1}{z}\Vmz\right)=iz\left[\pi,\left(\frac{1}{z}\Amz\right)\right]~,
\end{align}
so it makes sense to solve the equations in terms of $\Amz,\Vmz$ rather than $\Am,\Vm$.
Let us first define 
\begin{align}
Y^\pm_\mu \equiv  e^{\pm\frac{i}{2}\left(z^2-1\right)\pi} X_\mu e^{\mp \frac{i}{2}\left(z^2-1\right)\pi}~,
\end{align}
for an arbitrary algebra-valued, $z$-independent $X_{\mu}$. Then it is easily seen to satisfy
\begin{align}
\dz Y^\pm_\mu =\pm i z \Big[ \pi , Y^\pm_\mu \Big]~,
\end{align}
and
\begin{align}\label{eq:Y's}
Y_\mu^\pm(z=1)=X_\mu~.
\end{align}
We immediately see that 
\begin{align}\label{eq:Amz_Vmz_solution}
\Amz&=\frac{z}{2}\left(Y_\mu^+ + Y_\mu^-\right)\nn\\
 \Vmz&=\frac{z}{2}\left(Y_\mu^+ - Y_\mu^-\right)
\end{align}
solves the equations of motion \eqref{eq:Amz_Vmz_EOM}, and
\begin{align}
\Vmz(z=1)=0~,~~~\Amz(z=1)=X_\mu~~.
\end{align}
More explicitly it is,
\begin{align}\label{eq:AmzVmz}
\Amz&=\frac{z}{2}\left( e^{+\frac{i}{2}\left(z^2-1\right)\pi} X_\mu e^{- \frac{i}{2}\left(z^2-1\right)\pi} + e^{-\frac{i}{2}\left(z^2-1\right)\pi} X_\mu e^{+ \frac{i}{2}\left(z^2-1\right)\pi}\right)~, \nn\\
\Vmz&=\frac{z}{2}\left(e^{+\frac{i}{2}\left(z^2-1\right)\pi} X_\mu  e^{- \frac{i}{2}\left(z^2-1\right)\pi} - e^{-\frac{i}{2}\left(z^2-1\right)\pi} X_\mu e^{+ \frac{i}{2}\left(z^2-1\right)\pi}\right)~.
\end{align}
  The two-derivative on-shell action then becomes,
\begin{align}
\cS_{5d}&=\frac{1}{g_4^2}\int d^4x\int_0^1\frac{dz}{z} \ \Tr\Big\{ \Vmz
V^{\mu}_{~~z}+\Amz A^{\mu}_{~~\!\!z} \Big\} =\frac{1}{2g_4^2}\int d^4x \Tr \Big\{X_{\mu}X^{\mu}\Big\} ~.
\end{align}
However, note that  $X_\mu$ is arbitrary in \eqref{eq:Y's}\eqref{eq:Amz_Vmz_solution}.\footnote{In order to have $\pi$-even $\Vmz$ and $\pi$-odd $\Amz$, we need $X_\mu$ to be $\pi$-odd.} 
In fact, in terms of $\Amz$ and $\Vmz$ we are missing one boundary condition, we simply cannot express $\Am(z=1)=0$, so we cannot fix $X_\mu$ either without actually going back to solving the equations of motion in terms of $\Am$ and $\Vm$ (which then goes back to what we have done so far, at finite order in pions).   Instead, we can compare with the unique results at this order, dictated by the symmetries,
\begin{align}
\cS_{\pi,2}= \frac{f_{\pi}^2}{4} \int d^4x ~\Tr\Big\{\dm \Ud \dmu U\Big\} = f_{\pi}^2 \int d^4x  ~ \Tr\Big\{ \cD_\mu \pi \cD^\mu \pi\Big\}~,
\end{align}
where the pion's covariant derivative is defined through
\begin{align}
\cD_\mu \pi=&-\frac{i}{2}\left(\xi^\dagger \dm \xi - \xi \dm \xi^\dagger\right)\nn~,
\end{align}
and 
\begin{align}
\xi=\sqrt{U}=e^{i\pi/2}~.
\end{align}
Thus, we identify
\begin{align}\label{eq:Xmu}
X_\mu = 2\cD_\mu \pi=
\dm\pi - \frac{1}{24}\Big[\pi,\big[\pi,\dm\pi\big]\Big]+\cO(\pi^5)~.
\end{align}
Finally, we can plug \eqref{eq:Xmu} into \eqref{eq:AmzVmz} and expand, to get
\begin{align}
\Amz
&=z\dm\pi -\frac{1}{24}z\left(3z^4-6z^2+4\right)\Big[\pi,\big[\pi,\dm\pi\big]\Big]+\cO(\pi^5,\d^3)~,\nn\\
\Vmz
&= \frac{i}{2} z (z^2-1)\big[\pi,\dm\pi\big] +\cO(\pi^4,\d^3)~,
\end{align}
consistently with our perturbative results.

\section{5d Green's Functions and Feynman Rules}\label{app:Feynman}
We collect here the Green's functions of the propagating 5d fields in the gauge \eqref{eq:gauge_fixing}\eqref{eq:our_pion}, used throughout the paper. The vector and axial vector propagators are written as piecewise functions along
the $z$ direction. This form lends itself most straightforwardly to obtaining results to arbitrary order in momentum, and it is simple to verify that the poles
of these propagators correspond to the masses of the vector and axial-vector states predicted by the Hirn-Sanz model.
We begin by finding the Green's functions of the Abelian part of the
equations of motion.
\subsection{$A_z$ Wavefunction and Pion Propagator}
As described in the body of the text, we can choose a gauge where
\begin{align}
A_z(x,z)=z\pi(x)~,
\end{align}
to all orders in $\pi$. (As usual, we set $L=1$ for convenience
and restore it using dimensional analysis in the final result.) $A_z$ does not propagate in 5d, and does not run on
internal legs.

The quadratic order action for pions comes from
\begin{align}
S_{\pi^2} = -2\frac{R}{4g_5^2}\int  dz d^4x \Tr\Big\{\sqrt{g}A_{MN}^a A^{aMN} \Big\}\supset \frac{R}{2g_5^2}\int  d^4x \ \Tr\Big\{ \d_\mu\pi^a \d^\mu\pi^a\Big\}~,
\end{align}
where we used $A_z = z\pi $ and integrated over $z$. The pion propagator in momentum space thus takes the form,
\begin{align}
\big\langle\pi^a(k)\pi^b(0)\big\rangle =\left(\frac{g_5^2}{R}\right) ~ \frac{i}{k^2}\delta^{ab}~.
\end{align}

\subsection{$V_\mu$ Propagator}\label{app:GreenVmu}
In the $V_z=0$ gauge there are no ghosts in the 5d theory (and anyway we consider only tree-level diagrams) so
we need only consider the Green's function of the fields $V_\mu$.
The two-point function takes the form
\begin{align}\label{eq:Vprop}
\big\langle V_\mu^a(k,z) V_\nu^b(q,z')\big\rangle =
-i\frac{g_5^2}{2R} ~ (2\pi)^4\delta(k+q)\delta^{ab}~\left[G_V(k,z,z')\left(\eta_{\mu\nu}-\frac{k_\mu
k_\nu}{k^2} \right)+G_V(0,z,z')\frac{k_\mu k_\nu}{k^2}  \right]~,
\end{align}
where the $k$-momentum and zero-momentum Green's functions
satisfy
\begin{align}
z\d_z\frac{1}{z}\d_z G_V(k,z,z') +
k^2G_V(k,z,z')=&z
\delta(z-z')~,\nn\\
z\d_z\frac{1}{z}\d_z G_V(0,z,z')=&z \delta(z-z')~.
\end{align}
The vector mode  satisfies the boundary conditions
\begin{align}
\d_z G_V(1,z')=G_V(0,z')=0~.
\end{align}
One can solve for the Green's
function piecewise and show that
\begin{align}
G_{V}(k,z,z')&=-\frac{\pi z z'}{2}\left(\frac{Y_0(k)}{J_0(k)}J_1(kz)J_1(kz')-J_1(kz_-)Y_1(kz_+)\right)~,
\end{align}
when $z_+$ ($z_-$) is the larger (resp. smaller) of $z,z'$. In particular, we then have 
%\begin{align}
%z_+\equiv\left\{\begin{array}{cl}
%z' & \text{for} \  z'>z \\
%z & \text{for} \ z>z'
%\end{array} \right.
%\end{align}
\begin{align}
G_V(0,z,z')=-\frac{1}{2}z_-^2~,
\end{align}
and there is no pole at $k=0$. To second order in momentum,
\begin{align}
G_V(k,z,z')=-\frac{1}{2}z_-^2+k^2\frac{1}{16}\left(z_-^4-2z^2z'^2+4z^2z'^2\log(z_+) \right) +\cO(k^4)~.
\end{align}

\subsection{$A_\mu$ Propagator}\label{app:GreenAmu}
The story is almost identical for the $A_\mu$ propagator, except that the boundary condition at $z=1$ is now $A_\mu(z=1)=0$.
The two-point function takes the form
\begin{align}
\big\langle A_\mu^a(k,z) A_\nu^b(q,z')\big\rangle =
-i\frac{g_5^2}{2R} ~ (2\pi)^4\delta(k+q)\delta^{ab}~\left[G_A(k,z,z')\left(\eta_{\mu\nu}-\frac{k_\mu
k_\nu}{k^2} \right)+G_A(0,z,z')\frac{k_\mu k_\nu}{k^2}  \right]~,
\end{align}
where the $k$-momentum and zero-momentum Green's functions
satisfy the same equations of motion as those of the vector. (Note that there are no quadratic order cross terms between $A_\mu$ and $A_z$. This is a direct result of 
the gauge choice $A_z=z\pi$.)
The $A_\mu$ Green's function thus takes the form
\begin{align}
G_{A}(k,z,z')&=-\frac{\pi}{2}zz' \left(\frac{Y_1(k)}{J_1(k)}J_1(kz)J_1(kz')- J_1(kz_-)Y_1(kz_+)\right)~,
\end{align}
and
\begin{align}
G_{A}(0,z,z')&=-\frac{1}{2}z_-^2(1-z_+^2)~.
\end{align}

\subsection{Feynman Vertices in the Bulk }
Here we collect the Feynman vertices from the bulk action, written in momentum space along the flat spacetime directions, and position space for the radial direction. All momenta are assumed to be incoming.
Each $n$-leg vertex is accompanied by the integral
\begin{align}
\frac{R}{g_s^2}\int \frac{dz}{z} \frac{d^4k_1\dots d^4k_n}{(2\pi)^{4n}} (2\pi)^4\delta(k_1+k_2+\dots+k_n) ~.
\end{align}
When we compute the 4d off-shell action, the $A_z$ lines are
external only. They are only propagating (in 4d) when we
compute the S-matrix or the on-shell action. $V_\mu$ and $A_\mu$ lines are always
internal.

\begin{tikzpicture}
\path (180:2cm) node (Vp) {$V^c_\mu,k_3$}
    (60:2cm) node (Azk) {$\pi^a,k_1$}
    (300:2cm) node (Azq) {$\pi^b,k_2$}
    (2cm,0cm) node[anchor=west] (label) {$\sqrt{2}f^{abc}z^2\pi^a(k_1)\pi^b(k_2)(k_2-k_1)_\mu
$};
\draw[-,decorate,decoration=snake] (0,0) -- (Vp);
\draw[-] (0,0) -- (Azk);
\draw[-] (0,0) -- (Azq);
\end{tikzpicture}
\\
\\
\begin{tikzpicture}
\path (300:2cm) node (Vp) {$V^b_\mu,k_2$}
    (60:2cm) node (Ak) {$A^c_\nu,k_3$}
    (180:2cm) node (Azq) {$\pi^a,k_1$}
    (2.5cm,0) node[anchor=west] (label) {$-2\sqrt{2}if^{abc}z\pi^a(k_1)\eta_{\mu\nu}\d_z^{(V)}$};
\draw[-,decorate,decoration=snake] (0,0) -- (Vp);
\draw[-,decorate,decoration=snake] (0,0) -- (Ak);
\draw[-] (0,0) -- (Azq);
\end{tikzpicture}
\\
\\
\begin{tikzpicture}
\path (315:2cm) node (Vp) {$V^c_\mu,k_3$}
    (45:2cm) node (Vk) {$V^d_\nu,k_4$}
    (135:2cm) node (Azq) {$\pi^a,k_1$}
    (225:2cm) node (Azr) {$\pi^b,k_2$}
    (3cm,0) node[anchor=west] (label) {$i(f^{ace}f^{bde}+f^{adf}f^{bcf})z^2\pi^a(k_1)\pi^b(k_2)\eta_{\mu\nu}$};
\draw[-,decorate,decoration=snake] (0,0) -- (Vp);
\draw[-,decorate,decoration=snake] (0,0) -- (Vk); \draw[-] (0,0) --
(Azq); \draw[-] (0,0) -- (Azr);
\end{tikzpicture}
\\
\\
\begin{tikzpicture}
\path (315:2cm) node (Ap) {$A^a_\mu(p,z)$}
    (45:2cm) node (Ak) {$A^b_\nu(k,z)$}
    (135:2cm) node (Azq) {$A^c_z(q,z)$}
    (225:2cm) node (Azr) {$A^d_z(r,z)$}
    (2cm,0) node[anchor=west](label) {$\frac{iR}{g_5^2}(f^{acf}f^{bdf}+f^{adf}f^{bcf})\frac{z^2}{L^2}\pi^c(q)\pi^d(r)\eta_{\mu\nu}$};
\draw[-,decorate,decoration=snake] (0,0) -- (Ap);
\draw[-,decorate,decoration=snake] (0,0) -- (Ak); \draw[-] (0,0) --
(Azq); \draw[-] (0,0) -- (Azr);
\end{tikzpicture}
\\
\\
\begin{tikzpicture}
\path (180:2cm) node (Vp) {$V^a_\mu(p,z)$}
    (60:2cm) node (Vk) {$V^b_\nu(k,z)$}
    (300:2cm) node (Vq) {$V^c_\rho(q,z)$}
    (2cm,0) node[anchor=west] (label) {$\frac{\sqrt{2}R}{g_5^2}f^{abc}\left[(p_\rho-k_\rho)\eta_{\mu\nu}+(q_\nu-p_\nu)\eta_{\rho\mu}+(k_\mu-q_\mu)\eta_{\nu\rho}\right]$};
\draw[-,decorate,decoration=snake] (0,0) -- (Vp);
\draw[-,decorate,decoration=snake] (0,0) -- (Vk);
\draw[-,decorate,decoration=snake] (0,0) -- (Vq);
\end{tikzpicture}
\\
\\
\begin{tikzpicture}
\path (180:2cm) node (Vp) {$V^a_\mu(p,z)$}
    (60:2cm) node (Ak) {$A^b_\nu(k,z)$}
    (300:2cm) node (Aq) {$A^c_\rho(q,z)$}
    (2cm,0) node[anchor=west] (label) {-$\frac{\sqrt{2}R}{g_5^2}f^{abc}\left[(q_\mu-k_\mu)\eta_{\rho\nu}+(p_\nu-q_\nu)\eta_{\rho\mu}+(k_\rho-p_\rho)\eta_{\mu\nu}\right]$};
\draw[-,decorate,decoration=snake] (0,0) -- (Vp);
\draw[-,decorate,decoration=snake] (0,0) -- (Ak);
\draw[-,decorate,decoration=snake] (0,0) -- (Aq);
\end{tikzpicture}
\\
\begin{tikzpicture}
\path (45:2cm) node (Vp) {$V^a_\mu(p,z)$}
    (135:2cm) node (Vk) {$V^b_\nu(k,z)$}
    (225:2cm) node (Vq) {$V^c_\rho(q,z)$}
    (315:2cm) node (Vr) {$V^d_\sigma(r,z)$}
        (2cm,0cm) node[anchor=west,text width=10cm] (label)   {$-\frac{iR}{g_5^2}\left[ f^{abf}f^{cdf}(\eta_{\mu\rho}\eta_{\nu\sigma}-\eta_{\nu\rho}\eta_{\mu\sigma})\right.$ \\
        $\qquad\qquad+f^{acf}f^{bdf}(\eta_{\mu\nu}\eta_{\rho\sigma}-\eta_{\nu\rho}\eta_{\mu\sigma})$\\ $\left. \qquad\qquad+f^{bcf}f^{adf}(\eta_{\mu\nu}\eta_{\rho\sigma}-\eta_{\mu\rho}\eta_{\nu\sigma})\right]$};
\draw[-,decorate,decoration=snake] (0,0) -- (Vp);
\draw[-,decorate,decoration=snake] (0,0) -- (Vk);
\draw[-,decorate,decoration=snake] (0,0) -- (Vq);
\draw[-,decorate,decoration=snake] (0,0) -- (Vr);
\end{tikzpicture}
\\
\\
\begin{tikzpicture}
\path (45:2cm) node (Ap) {$A^a_\mu(p,z)$}
    (135:2cm) node (Ak) {$A^b_\nu(k,z)$}
    (225:2cm) node (Aq) {$A^c_\rho(q,z)$}
    (315:2cm) node (Ar) {$A^d_\sigma(r,z)$}
    (2cm,0cm) node[anchor=west,text width=10cm] (label)   {$-\frac{iR}{g_5^2}\left[ f^{abf}f^{cdf}(\eta_{\mu\rho}\eta_{\nu\sigma}-\eta_{\nu\rho}\eta_{\mu\sigma})\right.$ \\
        $\qquad\qquad+f^{acf}f^{bdf}(\eta_{\mu\nu}\eta_{\rho\sigma}-\eta_{\nu\rho}\eta_{\mu\sigma})$\\ $\left. \qquad\qquad+f^{bcf}f^{adf}(\eta_{\mu\nu}\eta_{\rho\sigma}-\eta_{\mu\rho}\eta_{\nu\sigma})\right]$};
        \draw[-,decorate,decoration=snake] (0,0) -- (Ap);
\draw[-,decorate,decoration=snake] (0,0) -- (Ak);
\draw[-,decorate,decoration=snake] (0,0) -- (Aq);
\draw[-,decorate,decoration=snake] (0,0) -- (Ar);
\end{tikzpicture}
\\
\\
\begin{tikzpicture}
\path (45:2cm) node (Vp) {$V^a_\mu(p,z)$}
    (135:2cm) node (Vk) {$V^b_\nu(k,z)$}
    (225:2cm) node (Aq) {$A^c_\rho(q,z)$}
    (315:2cm) node (Ar) {$A^d_\sigma(r,z)$}
    (2cm,0cm) node [anchor=west,text width=10cm] (label) {$-\frac{iR}{g_5^2}\left[ f^{abf}f^{cdf}(\eta_{\mu\rho}
\eta_{\nu\sigma}-\eta_{\nu\rho}\eta_{\mu\sigma})\right.$ \\ $\qquad\qquad+f^{acf}f^{bdf}(\eta_{\mu\nu}\eta_{\rho\sigma}-\eta_{\nu\rho}\eta_{\mu\sigma})$\\ $\left. \qquad\qquad+f^{bcf}f^{adf}(\eta_{\mu\nu}\eta_{\rho\sigma}-\eta_{\mu\rho}\eta_{\nu\sigma})\right]$};
    \draw[-,decorate,decoration=snake] (0,0) -- (Vp);
\draw[-,decorate,decoration=snake] (0,0) -- (Vk);
\draw[-,decorate,decoration=snake] (0,0) -- (Aq);
\draw[-,decorate,decoration=snake] (0,0) -- (Ar);
\end{tikzpicture}

\section{Six-pion interaction}\label{app:6pi}

We present here a more detailed derivation of the six-pion interaction term in the effective action.
The three Feynman diagrams which contribute to this interaction are shown in Figure \ref{fig:6pi}, labelled (1), (2), and (3). For each diagram, we have to sum over the possible
 external pion legs. Since not all of the $6!$ possible combinations are independent, we have to divide the net result in each case by a symmetry factor, as indicated
in each case below.

In Diagram (1), we must divide by a symmetry factor of $48$.
Noting that only the $\eta_{\mu\nu}$ pieces of the vector propagators contribute at this order in $\pi$ and after some manipulations, we find the
contribution of Diagram (1):
\begin{align}
S_{6,1}&=-\frac{R}{6g_5^2}\int \frac{\left(\prod_id^4k_i\right)\delta^{(4)}(\sum_i k_i)}{\left(2\pi\right)^{20}} \Tr \bigg\{ \Big[\pi(k_5),\big[\pi(k_1),\pi(k_2)\big]\Big] \Big[\pi(k_6),\big[\pi(k_3),\pi(k_4)\big]\Big]\bigg\}\times\nn\\
&\hspace{2.5cm} I_{6,1}(k_{12},k_{34},k_{56}) \Big[ (k_1+k_2)\cdot (k_5-k_6)(k_1-k_2)\cdot (k_3-k_4)\nonumber\\
&\hspace{5.5cm} +(k_3+k_4)\cdot (k_1-k_2)(k_3-k_4)\cdot (k_5-k_6)\nonumber\\
&\hspace{5.5cm}+ (k_5+k_6)\cdot (k_3-k_4)(k_5-k_6)\cdot (k_1-k_2)\Big]\nonumber ~,
\end{align}
where the integral $I_{6,1}$  (and $I_{6,2}$, $I_{6,3}$ for the other diagrams) are defined in equation \ref{eq:6piints}.

In Diagram (2), the symmetry factor is $8$. Both the $\eta_{\mu\nu}$ and $k_\mu k_\nu$ tensor structures from the axial propagator contribute, while we can again neglect the latter in the vector
propagators. The Diagram (2) contribution thus amounts to
\begin{align}
S_{6,2}=&-\frac{R}{g_5^2}\int \frac{\left(\prod_id^4k_i\right)\delta^{(4)}(\sum_i k_i)}{\left(2\pi\right)^{20}} \Tr \bigg\{ \Big[\pi(k_5),\big[\pi(k_1),\pi(k_2)\big]\Big] \Big[\pi(k_6),\big[\pi(k_3),\pi(k_4)\big]\Big]\bigg\}\times\nn\\
&\hspace{2cm}\bigg[ (k_1-k_2)\cdot (k_3-k_4) I_{6,2}(k_{12},k_{34},|k_1+k_2+k_5|) \nn\\
&\hspace{2.3cm} - k_5\cdot (k_1-k_2)~ k_6\cdot (k_3-k_4) \frac{I_{6,2}(k_{12},k_{34},|k_1+k_2+k_5|)-I_{6,2}(k_{12},k_{34},0)}{(k_1+k_2+k_5)^2}\bigg]\nn~.\end{align}
Finally, Diagram (3) (with symmetry factor $16$) yields,
\begin{align}
S_{6,3}=-\frac{R}{4g_5^2}&\int \frac{\left(\prod_id^4k_i\right)\delta^{(4)}(\sum_i k_i)}{\left(2\pi\right)^{20}}  \Tr \bigg\{ \Big[\pi(k_5),\big[\pi(k_1),\pi(k_2)\big]\Big] \Big[\pi(k_6),\big[\pi(k_3),\pi(k_4)\big]\Big]\bigg\}\times\nn\\
&\hspace{1.5cm}I_{6,3}(k_{12},k_{34}) (k_1-k_2)\cdot(k_3-k_4)~.
\end{align}
Summing these three contributions yields the full 6-pion term in the off-shell effective action, equation (\ref{eq:fullS6}).

\section{Off-Shell Identifications}\label{app:off-shell}

We have repeatedly emphasized that the holographic pion that we have defined is not the same as the NG pion: the two are only expected to have some overlap. Accordingly,  the on-shell quantities need to agree. Indeed, we have successfully matched the on-shell effective action of our pion with that of Hirn and Sanz, defined for the NG pion, finding full agreement. On the other hand, we can also compare some \textit{off-shell} quantities in order to extract the precise relation between them (as in \eqref{eq:overlap}). In this Appendix we compare the two pions' off-shell effective Lagrangian, as well as their equations of motion, and find  the same relation in both cases. We also show that this is the relation that should have been anticipated a priori, upon identifying the NG pion  (that is, the pion defined in terms of the nonlinear realization of the symmetry) in the model under consideration.

Consider first the chiral Lagrangian. From 
subsection \ref{subsec:comparison} we read off (already using the right relations between the $L_i$'s),
\begin{align}\label{eq:LPi}
\cL_\Pi
&=\frac{1}{4}\Tr\left(\dm\Pi\dmu\Pi+\frac{1}{12f_\Pi^2}\big[\Pi,\dm \Pi\big] \big[\Pi, \dmu\Pi\big]+\frac{1}{360f_\Pi^4}\Big[\Pi,\big[\Pi,\dm\Pi\big]\Big]\Big[\Pi,\big[\Pi,\dmu\Pi\big]\Big]\right.\nn\\
&~\left.-\frac{4L_1}{f_\Pi^4}\big[\dm\Pi,\dn\Pi\big]\big[\dmu\Pi,\dnu\Pi\big]+...\right)~.
\end{align}
We compare this to  the off-shell Lagrangian derived holographically in section \ref{sec:diagrammatics},
\begin{align}
\cL_\pi&=\frac{1}{2g_4^2}\Tr \left(\dm \pi\dmu\pi +\frac{L^2}{12}\big[\pi,\dm\pi\big]\big[\pi,\dmu\pi\big]+\frac{L^4}{360}\Big[\pi,\big[\pi,\dm\pi\big]\Big]\Big[\pi,\big[\pi,\dmu\pi\big]\Big]\right.\nn\\
&~\left.+\frac{11L^4}{384}\big[\dm\pi,\dn\pi\big]\big[\dmu\pi,\dnu\pi\big]+...\right)~,
\end{align}
and rederive our previous results \eqref{eq:fpi}\eqref{eq:solution_Li's}. However, now we also find the off-shell relation between the pions (up to the  order in pions we consider), 
\begin{align}\label{eq:pions_relation}
\pi=c\Pi+O(\Pi^7,\d^2\Pi^5)~,
\end{align}
with $c=g_4/\sqrt{2}=1/(Lf_\Pi)$ as in \eqref{eq:fpi}. 

The same off-shell information can be derived by comparing the pions' equations of motion. In \eqref{eq:pi_eom(3,4)} we found the equation of motion for our pion,
\begin{align} \label{eq:our_pion_EOM}
\d^2\pi=\frac{L^2}{6}\Big[\dmu\pi,\big[\pi,\dm\pi\big]\Big]+\frac{11L^4}{192}\Big[\dnu\pi,\big[\dmu\pi,\dn\dm\pi\big]\Big]+O(\d^2\pi^5,\d^6\pi^3)~.
\end{align}
For the NG pion, the equations of motion follow directly from \eqref{eq:LPi},
\begin{align}
\d^2\Pi&=\frac{1}{6f_\Pi^2}\Big[\dmu\Pi,\big[\Pi,\dm\Pi\big]\Big]+\frac{8L_1}{f_\Pi^4}\Big[\dnu\Pi,\big[\dmu\Pi,\dm\dn\Pi\big]\Big]+O(\d^2\Pi^5,\d^6\Pi^3)~,
\end{align}
and a comparison with \eqref{eq:our_pion_EOM} immediately leads to the same relation \eqref{eq:pions_relation}.
It should be noted that, while here we have found a very simple relation between the holographic pion and the NG pion, in the general case the relation may be much more complicated, with infinite-order corrections. (The method described here can be used generically to derive this relation order by order.)  To understand our result, we can easily identify the NG pion in this model as the non-trivial holonomy along the radial direction \cite{Hirn:2005nr},\footnote{The factor of 2 is crucial for the correct transformation law under the broken flavor symmetry. }
\begin{align}\label{eq:holonomy}
U(x)\equiv \exp\left(i\frac{\Pi(x)}{f_\Pi}\right)=P\left\{\exp\left(2\int_0^L dz \Az(x,z)\right)\right\}~,
\end{align} 
which can be shown to exhibit the right transformation law under the full flavor symmetry, $U\rightarrow L U R^\dagger$.
After the (on-shell) gauge fixing \eqref{eq:Az_gauge_fix}, the identification  \eqref{eq:holonomy} yields
\begin{align}
\pi(x)=\frac{\Pi(x)}{Lf_{\Pi}}~,
\end{align}
in agreement with the result above.

\bibliographystyle{JHEP}
\bibliography{domokosfieldarxiv}

\end{document}